\def\Roman#1{\uppercase\expandafter{\romannumeral#1}}
\documentclass[a4paper,12pt]{article}
\usepackage{cite}
\usepackage{amsmath}
\usepackage{amssymb,amsfonts}
\usepackage[mathscr]{eucal}
\usepackage{mathrsfs}
\usepackage[utf8x]{inputenc}

\title{Reduction  of path integrals  for interacting systems: The case of using dependent coordinates in the description of reduced motion on the orbit space}

\author{S. N. Storchak\footnote{E-mail adress: storchak@ihep.ru}\\
\small{ Institute for High Energy Physics,}\\
\small{ NRC ``Kurchatov Institute,''}\\
 \small{Protvino, 142284, Russian Federation}}

\begin{document}

  \maketitle

\begin{abstract}
We consider a reduction procedure in Wiener-type path integral for a finite-dimensional mechanical system with a symmetry representing the motion of two interacting scalar particles on a manifold that is the product of the total space of the principal fiber bundle and a vector space. By analogy with what is done in gauge theories, the local description of the reduced motion on the space of orbits is carried out using dependent coordinates.
The factorization of the measure in the path integral, which is necessary for the reduction, is based on the application of the stochastic differential equation of the optimal nonlinear filtering   from the theory of stochastic processes. The non-invariance of the measure in the path integral under the reduction is shown.  The integrand  of the path integral reduction  Jacobian is generated by the projection of the  mean curvature vector field of the orbit onto the submanifold,  which is used to determine the adapted coordinates in the principal fiber bundle  associated with the  problem under study. 
\end{abstract}

\section{Introduction}
Dynamical systems with symmetry are characterized by a deep relationship with geometry.
These include, for example, dynamical systems with gauge symmetries, especially those studied in gauge field theories, where the Yang-Mills fields are represented by connections on the principal fiber bundles.

In gauge theories, the main problem is to describe quantum evolution in terms of gauge invariant variables. This is the same as describing the evolution given on the space of orbits of the group action. 
A satisfactory solution has not yet been found for this problem.
At present,  in the path integral quantization of the gauge  systems, an approach based on the Faddeev-Popov mechanism \cite{Faddeev} is used. In this approach, the additional restrictions are imposed on the gauge fields, and therefore  we are forced to describe the quantum dynamics on the orbit space in terms of constrained (or dependent) variables.

However, the transition to constrained gauge fields, carried out in path integrals, looks like a heuristic  method of the path integral transformations.
To clarify the question of the correctness of such a transition, we studied how this happens in path integrals for finite-dimensional dynamical systems with symmetry\cite{Storchak_2,Storchak_2009}.
 This was done because for finite-dimensional Wiener-type path integrals, unlike of path integrals used in gauge theories, the integration measure can  be rigorously determined.

 The model system we studied was a mechanical system describing the motion of a scalar particle on a finite-dimensional manifold, on which the free isometric action of a compact semisimple Lie group is given.
The dynamics of this system is largely due to the presence of the principal fiber bundle associated with the considered dynamical system. The original configuration space of the system  can be viewed as a total space of this bundle. Therefore, as coordinates in the original configuration space, one can use the coordinates that are usually given in the principal fiber bundle.
The possibility of using a coordinate system including dependent coordinates follows from the existence of a local isomorphism between the original principal bundle and the trivial principal bundle whose base space is a local submanifold (`gauge surface')\cite{Mitter-Viallet}. 

However, in this approach, the problem of the global description of quantum evolution remains unsolved. 
From a geometric point of view, this is due to the fact that, in the general case, the principal bundles of gauge theories are nontrivial. 
Therefore, there are no global sections in the principal bundle, and we can only deal with local gauge fixing surfaces.
It also follows from this that a global description of evolution with the help of dependent variables defined on these surfaces is possible only when considering trivial principal bundles.

In our paper, the path integral reduction procedure  consists in transformation of  the original path integral, whose measure is generated by a stochastic process, to the path integral that determines the diffusion (or ``quantum evolution'') of the reduced dynamical system given on the orbit space  of the principal bundle.
 In this transformation, the main role is played by the factorization of the 
measure in the original path integral into the `group' measure (given in the space of paths on a group) and a measure defined in path space on a special surface of the principal bundle known as a `gauge fixing surface' (or `gauge surface').

In our papers\cite{Storchak_2, Storchak_1, Storchak_98} we used the definition of path integrals proposed by Daletskii and Belopolskaya in \cite{Dalecky_1,Dalecky_2}. According to their approach, the measure of path integrals is determined by those stochastic processes that are associated with local evolution semigroups.
 The global path integral is defined as the limit of these semigroups upon refining the subdivision of the time interval.
Transformations of path integrals (local evolution semigroups) can be obtained from transformations of local stochastic differential equations whose solutions are stochastic processes that generate measures in these path integrals.
In our works, the factorization of the measure in the path integrals was carried out using the stochastic differential equation of the optimal nonlinear filtering from the  stochastic process theory.
Note that a similar method for studying reduction in path integrals was used in \cite{Elworthy}.

 The path integral reduction leads to the integral relation between the original path integral and the path integral used to  describe the ``quantum evolution'' on the orbit space of the principal fiber bundle.
We also note that the Hamilton operator of the dynamical system obtained as a result of the reduction of the path integral has an additional potential term, which is  the integrand of the path integral reduction Jacobian.

In our articles \cite{Storchak_2019, Storchak_2020}, the path integral reduction method was applied to a simple model system describing the interaction of two scalar particles that are given on a manifold represented by the product of two manifolds:  a smooth compact Riemannian manifold and a vector space.  
The product manifold was endowed with a free proper isometric action of a semisimple compact unimodular Lie group.
 Such a group action gives rise to a principal fiber  bundle, so that the original configuration space, the product manifold, can be regarded as the total space of this bundle.
In the papers cited, we considered the case where the gauge surface (used to obtain a local section of the principal fiber bundle) has a parametric representation and, therefore, it is possible to define invariant variables that can serve as coordinates in the orbit space of this bundle.

The purpose of the present paper is to consider the case of the path integral reduction when, as in gauge theories, dependent variables are used to describe the motion on the orbit space. 
It should be noted that the geometric properties of the considered mechanical system are similar to the properties of an infinite-dimensional dynamical system describing the interaction of the Yang-Mills gauge field with a scalar field. 
This makes it possible to use the mechanical system under study as a finite-dimensional model for such field theories.

The paper is organizad as follows. In Section 2 we  give basic definitions that are in fact the same as those from our previous articles.  In Section 3 we briefly describe the geometry of the problem  and discuss the introduction of the adapted coordinates. Some of the information presented in this section was also borrowed from our previous work.   In Sections 4  we define the drift coefficients  of our stochastic differential equations. Section 5 deals with the factorization of the measure in the path integral and the calculation of the reduction Jacobian for the case of the reduction onto a zero-momentum level.  In  Conclusion, the details of the obtained result  are discussed.
 Some relationships between Christoffel symbols used in the article are given in   Appendix.

\section{Definitions}

In this article, we  consider  the  path integrals that are used to represent the solution  of the backward Kolmogorov equation which is given on a smooth compact Riemannian manifold $ \tilde {\mathcal P} = \mathcal P \times \mathcal V $:
\begin{equation}
\left\{
\begin{array}{l}
\displaystyle
\left(
\frac \partial {\partial t_a}+\frac 12\mu ^2\kappa \bigl[\triangle
_{\cal P}(p_a)+\triangle
_{\cal V}(v_a)\bigr]+\frac
1{\mu ^2\kappa m}V(p_a,v_a)\right){\psi}_{t_b} (p_a,t_a)=0,\\
{\psi}_{t_b} (p_b,v_b,t_b)=\phi _0(p_b,v_b),
\qquad\qquad\qquad\qquad\qquad (t_{b}>t_{a}),
\end{array}\right.
\label{1}
\end{equation}
where $\mu ^2=\frac \hbar m$ , $\kappa $  is a real positive
parameter,  $V(p,f)$ is the group-invariant potential term:
 $V(pg,g^{-1}v)=V(p,v)$, $g\in \mathcal G$, 
$\triangle _{\cal P}$ is the Laplace--Beltrami operator on the Riemannian
manifold $\cal P$. The scalar Laplacian $\triangle
_{\cal V}$ acts on the space of functions defined on  the finite-dimensional vector space $\cal V$. In a chart $(U_{\cal P}\times U_{\cal V},{\varphi}^{\tilde{\mathcal P}} )$, ${\varphi}^{\tilde{\mathcal P}}=(\varphi^A,\varphi^a)$, from the atlas of the manifold $\tilde{\mathcal P}$, where the point $(p,v)$ has the  coordinates $(Q^A,f^a)={\varphi}^{\tilde{\mathcal P}}(p,v)$, $\triangle _{\cal P}$ is given by the following expression\footnote{In  our formulas  we  assume that there is 
sum over the repeated indices. The indices denoted by the capital
letters  run from 1 to $n_{\cal P}=\dim \cal P$, and the indices of small Latin letters, except $i,j,k,l$, run from 1 to $n_{\cal V}=\dim \cal V$.}:
\begin{equation}
\triangle _{\cal P}(Q)=G^{-1/2}(Q)\frac \partial {\partial
Q^A}G^{AB}(Q)
G^{1/2}(Q)\frac\partial {\partial Q^B},
\label{2}
\end{equation}
where $G=det (G_{AB})$ and  $G_{AB}(Q)$ represents the Riemannian
metric on $\cal P$: $G_{AB}(Q)=G(\frac{\partial}{\partial Q^A},\frac{\partial}{\partial Q^B})$.

The coordinate expression of  the operator $\triangle_{\cal V}$ is  
\[
 \triangle_{\cal V}(f)=G^{ab}\frac{\partial}{\partial f^a\partial f^b},
\]
where  $G^{ab}$ is an inverse  matrix to the matrix $G_{ab}$ representing the metric  on $\mathcal V$  in the coordinate basis $\{\frac{\partial}{\partial f^a}\}$.
We assume that  the matrix $G_{ab}$ consists of fixed constant elements. 

To obtain the Schr\"odinger equation, one should perform the transition from the equation (\ref{1}) to the forward Kolmogorov equation and then set $\kappa =i$.

With the definition of the path integral from \cite{Dalecky_1}, the solution of the equation (\ref{1})
 can be written as 
\begin{eqnarray}
{\psi}_{t_b} (p_a,v_a,t_a)&=&{\rm E}\Bigl[\phi _0(\eta_1 (t_b),\eta_2(t_b))\exp \{\frac
1{\mu
^2\kappa m}\int_{t_a}^{t_b}V(\eta_1(u),\eta_2(u))du\}\Bigr]\nonumber\\
&=&\int_{\Omega _{-}}d\mu ^\eta (\omega )\phi _0(\eta (t_b))\exp
\{\ldots 
\},
\label{orig_path_int}
\end{eqnarray}
where ${\eta}(t)=(\eta_1(t),\eta_2(t))$ is a global stochastic process on a manifold 
$\tilde{\cal P}=\cal P\times \cal V$, 
the measure  ${\mu}^{\eta}$ in 
the space of paths $\Omega _{-}=\{\omega (t)=(\omega^1(t),\omega^2(t)): \omega^{1,2} (t_a)=0,
\eta_1
(t)=p_a+\omega^1 (t), \eta_2(t)=v_a+\omega^2(t)\}$ is determined by the probability distribution of the stochastic  process $\eta(t)$.

Note that  (\ref{orig_path_int}) is  a symbolical notation of the global semigroup, defined according to \cite{Dalecky_1, Dalecky_2} as a limit (under refinement of the  subdivision of the time interval)
of the superposition of the local semigroups: 
\begin{equation}
\!\psi _{t_b}(p_a,v_a,t_a)=
{\lim}_q \bigl[{\tilde U}_{\eta}(t_a,t_1)\cdot\ldots\cdot
{\tilde U}_{\eta}(t_{n-1},t_b)
\phi _0\bigr](p_a,v_a),
\label{6}
\end{equation}
where  each  local evolution semigroup
${\tilde U}_{\eta}$ acting in the space of functions on the manifold $\tilde {\mathcal P}$ is given by\footnote{In what follows, the potential term of the Hamilton operator, as being insignificant for our transformations of path integrals, will be temporarily omitted in the path integrals. In the final formulae, the potential term will be recovered.}
\begin{equation}
 {\tilde U}_{\eta}(s,t) \phi (p,v)={\rm E}_{s,p,v}[\phi (\eta_1
 (t),\eta_2(t))],\,\,\,\,\,\,
 s< t,\,\,\,\,\,\,\eta_1 (s)=p,\;\eta_2 (s)=v,
 \label{local_semigroup}
 \end{equation}
in which  ${\tilde U}_{\eta}(s,t) \phi (p,v)$ should be understand as $[{\tilde U}_{\eta}(s,t) \phi] (p,v)$. 

 These local semigroups are also expressed in terms of  path integrals with the integration measures determined by the local representatives
 $\varphi
 ^{\tilde{\cal P}}(\eta_t)
={\eta}^{\varphi^{\tilde{\cal P}}}(t)
\equiv\{\eta ^A_1 (t),\eta^a_2(t)\} \in R^{n_{\tilde{\cal P}}}$ of the global stochastic process $\eta(t)$.

The local processes     $\{\eta_1^A(t),\eta_2^a(t)\}$  are  solutions of the  stochasic differential equations 
\begin{align}
d\eta_1^A(t)&=\frac12\mu ^2\kappa G^{-1/2}\frac \partial {\partial
Q^B}(G^{1/2}G^{AB})dt+\mu \sqrt{\kappa }{\mathscr X}_{\bar{M}}^A(\eta_1
(t))dw^{
\bar{M}}(t),
\label{eta_1}\\
 d\eta_2^a(t)&=\mu \sqrt{\kappa }{\mathscr X}_{\bar{b}}^a
dw^{
\bar{b}}(t),
\label{eta_2}
\end{align}
in which the stochastic differentials are taken in the sence of It\^o
and where  ${\mathscr X}_{\bar{M}}^A$ and ${\mathscr X}_{\bar{b}}^a$ are  defined  by the local equalities
$\sum^{n_{\mathcal P}}_{\bar{\scriptscriptstyle K}\scriptscriptstyle =1}{\mathscr X
}_{\bar{K}}^A{\mathscr X}_{\bar{K}}^B=G^{AB}$ and  $\sum^{n_{\mathcal V}}_{\bar{\scriptscriptstyle b}\scriptscriptstyle =1}{\mathscr X
}_{\bar{b}}^a{\mathscr X}_{\bar{b}}^c=G^{ac}$, $dw^{\bar{M}}(t)$ and $dw^{\bar{b}}(t)$ are the independent Wiener processes.
(Here and further  we  denote the Euclidean
indices by over-barred indices.) Also note that equations (\ref{eta_1}) and (\ref{eta_2}) are the Stratonovich equations.

Note that with the local processes $\{\eta_1^A(t),\eta_2^a(t)\}$ one can rewrite (\ref{local_semigroup}) as 
\begin{equation}
 {\tilde U}_{\eta}(s,t) \phi (p,v)={\rm E}_{s,\varphi^{\tilde{\cal P}}(p,v)}[\phi (\,({\varphi}^{\tilde{\cal P}})^{-1}(\eta^{\varphi^{\tilde{\cal P}}}(t))\,)],\,\,\,\,\,\,
 {\eta}^{\varphi^{\tilde{\cal P}}}(s)=\varphi^{\tilde{\cal P}}(p,v).
 \label{local_semigroup_on_chart}
 \end{equation}

 The fundamental solution, also known as the Green's function 
$G_{\tilde{\cal P}}$,
is the kernel of the  semigroup, with the help of  which  the left-hand side of (\ref{orig_path_int}) is determined as follows:
\[
\psi_{t_b} (p_a,v_a,t_a)=\int G_{\tilde{\cal P}}(p_b,v_b,t_b;p_a,v_a,t_a)\phi _0(p_b,v_b)dv_{\tilde{\cal P}}(p_b,v_b),
\]
where $dv_{\tilde{\cal P}}(p,v)$ is a volume element on the manifold $\tilde{\cal P}$.
Note that in this formula the integration is carried out using a partition of unity subordinated to the local finite covering of the manifold $\tilde {\cal P}$.

The probabilistic representation for the kernel $ G_{\tilde{\cal P}} $ can be formally obtained by replacing $ \varphi _0 $ in 
(\ref{orig_path_int}) with the delta-function.
This means that the measure of the path integral for the Green's function $ G_{\tilde{\cal P}} $ will be defined in the space of paths with both ends fixed. 
The same can be done more correctly by considering functions approximating the delta function (instead of $\varphi_0$) and then taking the appropriate limit. 

The fundamental solution to the equation (\ref{1}) also satisfies the forward Kolmogorov equation.
As for the existence of a fundamental solution to the equation (\ref{1}), this depends on the smoothness constraints imposed on the coefficients of the equation, and in this article we assume that in our equation the coefficients are chosen in accordance with these requirements.

\section{Adapted coordinates in the principal fiber bundle}
Our main assumption in the article is that on the Riemannian manifold $ \tilde{\mathcal P} = \mathcal P \times \mathcal V $, which is the configuration space of our original mechanical system, there is a smooth isometric free and proper action of the compact semisimple Lie group $\mathcal G$. 
We consider the right action of the group by which
  the point $(p, v)$ is mapped to the point $ (\tilde p, \tilde v)$ so that
$ (\tilde p, \tilde v) = (p, v) g = (pg, g^{- 1} v) $.
If the point $(p,v)$ has the local coordinates $(Q^A,f^a)$ ( in the coordinate basis $(\frac{\partial}{\partial Q^A},\frac{\partial}{\partial f^a}$)), then this action in coordinates can be  written  as 
\[
 {\tilde Q}^A=F^A(Q,g),\;\;\;\;{\tilde f}^b=\bar D^b_a(g)f^a,
\]
where $\bar D^b_a(g)\equiv D^b_a(g^{-1})$,
and by $D^b_a(g)$ we denote the matrix of  the finite-dimensional representation of the group $\mathcal G$
acting in the vector space $\mathcal V$.

For the right action, we have the  compatibility relation 
\[
 F(F(Q,g_1),g_2)=F(Q,\rm \hat {\Phi}(g_1,g_2)),
\]
where the function $\rm\hat {\Phi}$  determines the group multiplication law in the space of the group parameters.

 The  manifold $\tilde{\mathcal P}$  is equipped with the following Riemannian metric:
\begin{equation}
 ds^2=G_{AB}(Q)dQ^AdQ^B+G_{ab}\,df^adf^b.
\label{metr_orig}
\end{equation}
We note, that the components of the metric are not arbitrary, but must satisfy certain relations due to the isometric action of the group $\mathcal G$ on $\tilde{\mathcal P}$.
 
In  particularly, 
\begin{equation}
 G_{AB}(Q)=G_{DC}(F(Q,g))F^D_A(Q,g)F^C_B(Q,g),
\label{relat_G_AB}
\end{equation}
with $ F^B_A(Q,g)\equiv \partial F^B(Q,g)/\partial Q^A$, and
\begin{equation}
 G_{pq}=G_{ab}\bar D^a_p(g)\bar D^b_q(g).
\label{relat_g_ab}
\end{equation}
The last relation can be derived from the linear isometrical action of the group $\mathcal G$ in the vector space $\mathcal V$.

The Killing vector fields $K_{\alpha}$, $\alpha=1,...,n_{\mathcal G}, n_{\mathcal G}=\dim \mathcal G$, of the manifold $\tilde{\mathcal P}$ with the metric (\ref{metr_orig})  are expressed in terms  of local coordinates $(Q^A,f^b)$ as follows:
\[
 K_{\alpha}=K^A_{\alpha}(Q)\partial / \partial Q^A+K^b_{\alpha}(f)\partial/\partial f^b,
\]
where 
$K^A_{\alpha}(Q)=\partial {\tilde Q}^A/\partial a^{\alpha}|_{a=e}$ and
$$K^b_{\alpha}(f)=\partial {\tilde f}^b/\partial a^{\alpha}|_{a=e}=\partial {\bar D}^b_c(a)/\partial a^{\alpha}|_{a=e}f^c\equiv({\bar J}_{\alpha})^b_c f^c.$$  
  The generators ${\bar J}_{\alpha}$ of the representation ${\bar D}^b_c(a)$ satisfy the  commutation relation 
$[{\bar J}_{\alpha},{\bar J}_{\beta}]={\bar c}^{\gamma}_{\alpha \beta}{\bar J}_{\gamma}$, where the structure constants
${\bar c}^{\gamma}_{\alpha \beta}=-{c}^{\gamma}_{\alpha \beta}$.

As it follows from the general theory (see, for example, \cite{AbrMarsd})
the  action of the group $ \mathcal G $ on the manifold $ \tilde {\mathcal P} $ leads to the principal fiber bundle $ \pi ': \mathcal P \times \mathcal V \to \mathcal P \times_{\mathcal G} \mathcal V $.\footnote{$\pi':(p,v)\to [p,v]$, where  $[p,v]$ is the equivalence class formed  by the  relation   $(p,v)\sim (pg,g^{-1}v).$} 
We  denote this principal bundle as $\rm P(\tilde{\mathcal M},\mathcal G)$, where the orbit space $\tilde{\mathcal M}=\mathcal P\times _{\mathcal G}\mathcal V$ is   the base space of the bundle.

The fact that the principal bundle is related to our problem means that the original manifold $\tilde{\mathcal P}$ can be considered as the total space of this  bundle.
It follows that we can express  the local coordinates $(Q^A, f^a)$ of the point $(p,v)\in \tilde{\mathcal P}$ in terms of the coordinates defined in the principal fiber bundle.

 As the coordinates in the principal  bundle, we will use the adapted coordinates. This choice is due to the fact that the quantization of gauge fields is mainly carried out using such coordinates \cite{Creutz,Huffel-Kelnhofer,Kelnhofer_2}.

Adapted coordinates
are defined with the help of `gauge fixing surfaces', which are local submanifolds $\tilde{\Sigma}_i$ in the total space of the principal bundle. 
There is a certain correspondence between these local submanifolds and the open neighborhoods $\tilde{U}_i$ from the atlas of the manifold defined on $\tilde {\mathcal M}$. In addition, $\tilde{\Sigma}_i$ must have a transversal intersection with the group orbits.

In our case, to determine $\tilde{\Sigma}_i$ in the principal fiber bundle $\rm P(\tilde {\mathcal M}, \mathcal G)$, we  use local surfaces $\Sigma_i$ in the total space $\mathcal P $ of the principal bundle $\rm P(\mathcal M, \mathcal G)$ with the base space $\mathcal M=\mathcal P/\mathcal G$.  In fact, the local surfaces $\Sigma_i$ serve to define local sections of this bundle. On the other hand, if the sections are given, then the images of these sections are the local surfaces $\Sigma_i$.

We assume that  local surfaces $\tilde{\Sigma}_i$ form a global surface $\tilde{\Sigma}$ in the original manifold $\tilde{\cal P}$ or, which is the same, that there exists a global section in the principal fiber bundle $\rm P(\tilde {\mathcal M}, \mathcal G)$. This implies that our principal fiber bundle  is trivial  and hence $\tilde{\cal P}\sim\tilde{\mathcal M}\times \mathcal G$. 
Such a case corresponds to what we have in gauge theories, where in practice one has to deal with trivial principal bundles.
Note also that triviality of $\rm P(\tilde {\mathcal M}, \mathcal G)$ is interrelated with the triviality of the principal fiber bundle $\rm P( {\mathcal M}, \mathcal G)$.

Let's consider how we can apply the general scheme of introducing adapted coordinates in our case.
The  submanifold $\Sigma $ of $\mathcal P$ in this approach is given by the system of equations
$\chi^{\alpha}=0,\,\alpha=1,...,n_{\mathcal G}$.  And the points belonging to $\Sigma$ have coordinates $Q^{\ast A}$ for which $ \chi^{\alpha}(Q^{\ast A})=0$. For this reason, the coordinates $Q^{\ast A}$ are called dependent coordinates.

The group coordinates $a^{\alpha}(Q)$ of a point $p\in \mathcal P$ are defined by the solution of the following equation:
\[
 \chi^{\alpha}(F^A(Q,a^{-1}(Q)))=0.
\]
This means that
\[
 Q^{\ast A}=F^A(Q,a^{-1}(Q)).
\]
That is, the group element $g^{-1}(p)$ with the coordinates $a^{-1}(Q)$ carries the point $p$ to the submanifold  $\Sigma$. 

Therefore, if we are given the (global) gauge fixing surface $\Sigma$ in the principal fiber bundle $\pi:\cal P\to\cal M$, then the section of the bundle can be defined as follows. Restricting the projection $\pi$ to $\Sigma$, we have 
$\pi|_{\Sigma}:U_{\Sigma}\to U_{\cal M}$ for local neighborhoods on $\Sigma$ and $ \cal M $. The section ${\sigma}_{\Sigma}$ is defined as a  map ${\sigma}_{\Sigma}:U_{\cal M}\to U_{\Sigma}$ such that $\pi|_{\Sigma}\circ{\sigma}_{\Sigma}={\rm id}_{\pi|_{\Sigma}}$. It follows from the above coordinate transformations that ${\sigma}_{\Sigma}([p])=p\,g^{-1}(p)$.

Using ${\sigma}_{\Sigma}$ one can define the section ${\sigma}_{\tilde\Sigma}$ of the  principal fiber bundle $\rm P(\tilde {\mathcal M}, \mathcal G)$ as the map ${\sigma}_{\tilde\Sigma}:\mathcal P \times_{\mathcal G} \mathcal V\to \mathcal P \times \mathcal V $. This section is defined as  ${\sigma}_{\tilde\Sigma}([p,v])=({\sigma}_{\Sigma}([p]),g(p) v)=(p\,g^{-1}(p),g(p)v)=(p,v)g^{-1}(p).$ The gauge fixing surface $\tilde{\Sigma}$ in $\mathcal P\times \mathcal V$ is the image of the mapping given by ${\sigma}_{\tilde\Sigma}$. \cite{Huffel-Kelnhofer}
 
 Having obtained $\tilde{\Sigma}$, we can define the principal  bundle $\pi'_{(\tilde{\Sigma})}:\mathcal P\times \mathcal V\to
{\tilde\Sigma}$ with $\pi'_{(\tilde{\Sigma})}={\sigma}_{\tilde\Sigma}\circ\pi'$. This principal bundle isomorphic to $\rm P(\tilde {\mathcal M}, \mathcal G)$. 

The group element $g^{-1}(p)$ obtained above (participating in the transition (along the orbit) from the point $p\in \mathcal P$ to the corresponding point given on the gauge fixing surface $\Sigma$ in the total space of the principal bundle ${\rm P}(\mathcal M,\mathcal G)$) is also used to determine adapted coordinates on the manifold $\tilde{\cal P}$.

 The coordinate functions $\tilde \varphi^{-1}$ that make the transition to adapted coordinates  are  as follows:
\[
 \tilde \varphi^{-1} :(Q^A,f^b)\to (Q^{\ast}{}^A(Q),\tilde f^b(Q),a^{\alpha}(Q)\,),
\]
where
\[
\tilde f^b(Q) = D^b_c(a(Q))\,f^c,
\]       
($\bar D^b_c(a^{-1})\equiv D^b_c(a))$ and $a(Q)$ are the coordinates of the group element $g(p)$).

The inverse transformation is given by the coordinate functions $\tilde \varphi$: 
\[
 \tilde \varphi :(Q^{\ast}{}^B,\tilde f^b,a^{\alpha})\to (F^A(Q^{\ast},a), \bar D^c_b(a)\tilde f^b).
\]
These functions,  $\tilde \varphi$ and $\tilde \varphi^{-1} $, represent the bundle maps $\tilde \varphi : \tilde {\Sigma}\times\cal G \to {\mathcal P} \times \cal V$ and 
${\tilde \varphi^{-1}}: {\mathcal P} \times {\mathcal V} \to {\tilde\Sigma} \times {\mathcal G}$ for the isomorphic trivial principal bundles $\tilde {\Sigma}\times\mathcal G \to \tilde {\Sigma} $ and ${\rm P}(\tilde{\mathcal M},\mathcal G)$.

Note also that in the general case when ${\rm P}(\tilde{\mathcal M},\mathcal G)$ is nontrivial,   the trivial principal  bundles ${\tilde\Sigma}_i \times {\mathcal G} \to {\tilde\Sigma}_i$ are only locally isomorphic to the principal bundle ${\rm P}(\tilde{\mathcal M},\mathcal G)$\cite{Huffel-Kelnhofer, Kelnhofer_2},\cite{Mitter-Viallet}. In this case the adapted coordinates can only be used to describe the local motion on the orbit space $\tilde{\mathcal M}$.

Thus, we have defined  the special  bundle coordinates in the total space of
 the principal fiber bundle 
$\pi':\mathcal P\times \mathcal V\to \mathcal P\times_{\mathcal G} \mathcal V$. They are such that 
$$\tilde{\varphi}^{\tilde{\cal P}}(p,v)=(Q^{\ast}{}^A,\tilde f^b, a^{\alpha})\;\;\rm {where}\;\; \tilde{\varphi}^{\tilde{\cal P}}:= \tilde \varphi^{-1}\circ \varphi^{\tilde{\cal P}}.$$

Note that adapted coordinates can be obtained in the same way using a finite set of fixing surfaces $\tilde{\Sigma}_i$.
But in this case,  it is also required that adapted coordinates must be consistent to each other on the  overlapping of the charts.

Replacing coordinates  $(Q^A,f^a)$ of a point $(p,v)\in \mathcal P\times V$  with new coordinates $(Q^{\ast}{}^B, \tilde f^b, a^{\alpha})$, 
\begin{equation}
Q^A=F^A(Q^{\ast}{}^B,a^{\alpha}),\;\;\;f^a=\bar D^a_b(a)\tilde f^b,
\label{transf_coord}
\end{equation}
 leads to the following transformations of the local coordinate vector fields: 
\begin{eqnarray}
\displaystyle 
&&\!\!\!\!\!\!\!\!\frac{\partial}{\partial f^a}=D^b_a(a)\frac{\partial}{\partial {\tilde f}^b},
\nonumber\\
&&\!\!\!\!\!\!\!\!\frac{\partial}{\partial Q^B}=\frac{\partial Q^{\ast}{}^A}{ \partial Q^B}\frac{\partial}{\partial Q^{\ast}{}^A}+\frac{\partial a^{\alpha}}{\partial Q^B}\frac{\partial}{\partial a^{\alpha}}+\frac{\partial {\tilde f}^b}{\partial Q^B}\frac{\partial}{\partial {\tilde f}^b}
\nonumber\\
&&\!\!\!\!\!\!\!\!\!\!\!\!\!=\check F^C_B\Biggl(N^A_C(Q^{\ast})\frac{\partial}{\partial Q^{\ast}{}^A}+{\chi}^{\mu}_C({\Phi}^{-1})^{\beta}_{\mu}\bar{v}^{\alpha}_{\beta}(a)\frac{\partial}{\partial a^{\alpha}}-{\chi}^{\mu}_C({\Phi}^{-1})^{\nu}_{\mu}(\bar J_{\nu})^b_p\tilde f^p\frac{\partial}{\partial {\tilde f}^b}\Biggr)\!.
\label{vectfield}
\end{eqnarray}
Here $\check F^C_B\equiv F^C_B(F(Q^{\ast},a),a^{-1})$ is an inverse matrix to the matrix $F^A_B(Q^{\ast},a)$,
${\chi}^{\mu}_C\equiv \frac{\partial {\chi}^{\mu}(Q)}{\partial Q^C}|_{Q=Q^{\ast}}$, $({\Phi}^{-1})^{\beta}_{\mu}\equiv({\Phi}^{-1})^{\beta}_{\mu}(Q^{\ast})$ -- the matrix which is inverse to the Faddeev--Popov matrix
\[
 ({\Phi})^{\beta}_{\mu}(Q)=K^A_{\mu}(Q)\frac{\partial {\chi}^{\beta}(Q)}{\partial Q^A},
\]
the matrix $\bar{v}^{\alpha}_{\beta}(a)$ is inverse of the matrix $\bar{u}^{\alpha}_{\beta}(a)$.\footnote{$\det \bar{u}^{\alpha}_{\beta}(a)$ is   the density of the right-invariant measure  given on the group $\mathcal G$.}

The operator $N^A_C$, defined  as
\[
 N^A_C(Q)=\delta^A_C-K^A_{\alpha}(Q)({\Phi}^{-1})^{\alpha}_{\mu}(Q){\chi}^{\mu}_C(Q), 
\]
 is the projection operator ($N^A_BN^B_C=N^A_C$) onto planes perpendicular to the gauge orbits in $ {\rm P}(\mathcal M, \mathcal G)$, since $N^A_CK^C_{\alpha}=0$\cite{Creutz}.  $N^A_C(Q^{\ast})$ is the restriction of $N^A_C(Q)$ to the submanifold $ \Sigma $:
\[
 N^A_C(Q^{\ast})\equiv N^A_C(F(Q^{\ast},e)),\;\;\;N^A_C(Q^{\ast})=F^B_C(Q^{\ast},a)N^M_B(F(Q^{\ast},a))\check F_M^A(Q^{\ast},a),
\]
where $e$ is the unity element of the group.

As an operator, the vector field $\frac{\partial}{\partial Q^{\ast}{}^A}$ is determined  by means of the following rule:
\[
 \frac{\partial}{\partial Q^{\ast}{}^A }\varphi(Q^{\ast})=(P _\bot)^D_A(Q^{\ast})\frac{\partial \varphi(Q)}{\partial Q^D}\Bigl|_{Q=Q^{\ast}},
\]
where the projection operator $(P_\bot)^A_B$ onto the tangent plane to the submanifold $\Sigma$ is given by
\[
 (P_\bot)^A_B=\delta^A_B-\chi ^{\alpha}_{B}\,(\chi \chi ^{\top})^{-1}{}^{\beta}_{\alpha}\,(\chi ^{\top})^A_{\beta}. 
\]
In this formula, $(\chi ^{\top})^A_{\beta}$ is a transposed matrix to the matrix $\chi ^{\nu}_B$:
\[
 (\chi ^{\top})^A_{\mu}=G^{AB}{\gamma}_{\mu\nu}\chi ^{\nu}_B\;\;\;\;{\gamma}_{\mu\nu}=K^A_{\mu}G_{AB}K^B_{\nu}.
\]
Using the above explicit expressions for the projection operators, it is easy to get
the following  properties:
 \[
 (P_\bot)^A_BN^C_A=(P_\bot)^C_B,\;\;\;\;\;N^A_B(P_\bot)^C_A=N^C_B.
\]

In the new coordinate basis $\displaystyle(\partial/\partial Q{}^{\ast A},\partial/\partial \tilde f^a,\partial/\partial a^{\alpha})$
the metric (\ref{metr_orig}) of the original manifold $\mathcal P \times \mathcal V$ is written as follows:
\begin{equation}
\displaystyle
{\tilde G}_{\cal A\cal B}(Q{}^{\ast},\tilde f,a)=\!
\left(
\begin{array}{ccc}
 G_{CD}(P_{\bot})^C_A (P_{\bot})^D_B & 0 & G_{CD}(P_{\bot})^C_AK^D_{\nu}\bar u^{\nu}_{\alpha}\\
 0 & G_{ab} & G_{ap}K^p_{\nu}\bar u^{\nu}_{\alpha}\\
G_{BC}K^C_{\mu}\bar u^{\mu}_{\beta} & G_{bp}K^p_{\nu}\bar u^{\nu}_{\beta} & d_{\mu\nu}\bar u^{\mu}_{\alpha}\bar u^{\nu}_{\beta}\\
\end{array}
\right),
\label{metric2c}
\end{equation}
where $G_{CD}(Q{}^{\ast})\equiv G_{CD}(F(Q{}^{\ast},e))$:
\[
 G_{CD}(Q{}^{\ast})=F^M_C(Q{}^{\ast},a)F^N_D(Q{}^{\ast},a)G_{MN}(F(Q{}^{\ast},a)), 
\]
the projection operator $P_\bot$ and the components $K^A_{\mu}$ of the Killing vector fields  depend on $Q{}^{\ast}$, $\bar u^{\mu}_{\beta}=\bar u^{\mu}_{\beta}(a)$,  $K^p_{\nu}=K^p_{\nu}(\tilde f)$,  $d_{\mu\nu}(Q{}^{\ast},\tilde f)$
is the tensor used to determine the metric given on the orbits of the group action. Its components are given by the relation 
\begin{eqnarray*}
 d_{\mu\nu}(Q{}^{\ast},\tilde f)&=&K^A_{\mu}(Q{}^{\ast})G_{AB}(Q{}^{\ast})K^B_{\nu}(Q{}^{\ast})+K^a_{\mu}(\tilde f)G_{ab}K^b_{\nu}(\tilde f)
\nonumber\\
&\equiv&\gamma_{\mu \nu}(Q^{\ast})+\gamma'_{\mu \nu}(\tilde f).
\end{eqnarray*}

Also we note that when obtaining (\ref{metric2c}), the  following transformations were used:
$$df^a=\bar D^a_p(a)d{\tilde f}^p+\frac{\partial \bar D^a_p(a)}{\partial a^{\mu}}\tilde f^p da^{\mu}$$
and
\[
 \frac{\partial \bar D^a_p(a)}{\partial a^{\mu}}\tilde f^p=(\bar J_{\beta})^c_p \bar D^a_c(a){\bar u}^{\beta}_{\mu}(a)\tilde f^p=K^c_{\beta}(\tilde f)\bar D^a_c(a){\bar u}^{\beta}_{\mu}(a).
\]
The last equality is due to 
the identity $D^c_b(a)(\bar J_{\alpha})^b_p\bar D^p_e(a)={\rho}^{\beta}_{\alpha}(a)(\bar J_{\beta})^c_e$, in which ${\rho}^{\beta}_{\alpha}(a)=\bar u^{\beta}_{\gamma}(a)v^{\gamma}_{\alpha}(a)$ is the matrix of the adjoint representation of the group $\mathcal G$.

The pseudoinverse matrix ${\tilde G}^{\cal A\cal B}(Q{}^{\ast},\tilde f,a)$ to  the matrix (\ref{metric2c}) is given by the following expression:
\begin{equation}
\displaystyle
\left(
\begin{array}{ccc}
{G}^{EF}N_E^AN_F^B & -G^{EF}N^A_E{\Lambda}^{\nu}_FK^a_{\nu} &  G^{EF}N^A_E{\Lambda}^{\beta}_F\bar v^{\alpha}_{\beta}\\
-G^{EF}N^B_F{\Lambda}^{\nu}_EK^b_{\nu} & G^{ba}+G^{EF}{\Lambda}^{\nu}_E{\Lambda}^{\mu}_FK^b_{\nu}K^a_{\mu} & -G^{EF}{\Lambda}^{\nu}_E{\Lambda}^{\mu}_FK^b_{\nu}{\bar v}^{\alpha}_{\mu}\\
G^{EF}N^B_F{\Lambda}^{\mu}_E\bar v_{\mu}^{\beta} & -G^{EF}{\Lambda}^{\nu}_E{\Lambda}^{\mu}_FK^a_{\mu}{\bar v}^{\beta}_{\nu} &G^{EF}{\Lambda}^{\nu}_E{\Lambda}^{\mu}_F{\bar v}^{\alpha}_{\nu}{\bar v}^{\beta}_{\mu}  \\
\end{array}
\right).
\label{metric2b}
\end{equation}
Here  ${\Lambda}^{\nu}_E\equiv({\Phi}^{-1})^{\nu}_{\mu}(Q{}^{\ast}){\chi}^{\mu}_E(Q{}^{\ast})$.

The pseudoinversion of ${\tilde G}_{\cal A\cal B}$ means that
\[
\displaystyle
{\tilde G}^{\tilde{\cal A}\tilde{\cal D}}{\tilde G}_{\tilde{\cal D}\tilde{\cal B}}=
\left(
\begin{array}{ccc}
  (P_{\bot})^A_B & 0 & 0\\
 0 & {\delta}^a_b & 0\\
0 & 0 & {\delta}^{\alpha}_{\beta}\\
\end{array}
\right).
\]
  
The determinant of the matrix  (\ref{metric2c}) is defined as
$$\det {\tilde  G}_{{\cal A\cal B}}=(\det d_{\mu\nu})\,(\det {\bar u}^{\mu}_{\nu})^2\,H,\,\footnote{In what follows, $\det d_{\mu\nu}$ will be denoted by $d$.}
\;\;\;\;\;\rm{with}$$ 
\begin{eqnarray}
 H= \displaystyle
\det \left(
\begin{array}{cc}
(P_{\bot})^{A'}_A\tilde G^{\rm H}_{A'B'}(P_{\bot})^{B'}_B & (P_{\bot})^{A'}_A\tilde G^{\rm H}_{A' a}\\
(P_{\bot})^{B'}_B\tilde G^{\rm H}_{b B'} & \tilde G^{\rm H}_{ab}\\
\end{array}
\right),
\label{det}
\end{eqnarray}
where
${\tilde G}^{\rm H}_{AB}=G_{AB}-G_{AC}K^C_{\mu}d^{\mu\nu}K^D_{\nu}G_{DB},$
$\;\;\tilde G^{\rm H}_{Aa}=-G_{AB}K^B_{\mu}d^{\mu\nu}K^b_{\nu}G_{ba}$, $\;\tilde G^{\rm H}_{ba}=G_{ba}-G_{bc}K^c_{\mu}d^{\mu\nu}K_{\nu}^pG_{pa}$.

Note that $\det {\tilde  G}_{{\cal A\cal B}}$ does not vanish only on the surface $\tilde \Sigma $. On this surface $\det(P_{\bot})^{A}_B$ is equal to unity.

It is also worth noting that  the matrix (\ref{metric2c}), representing the metric on  $\mathcal P \times \mathcal V$, can also be written in terms of the components 
of the  mechanical connection  existing  in the principal fiber bundle $\rm P(\tilde{\mathcal M},\mathcal G)$. This connection is given by a Lie-algebra valued one-form, which in local coordinates is represented as
$$\omega^{\alpha}={\mathscr A}^{\alpha}_D\,dQ^{\ast}{}^D+{\mathscr A}^{\alpha}_pd\tilde f^p+u^{\alpha}_{\mu}da^{\mu},$$
where
 $${\mathscr A}^{\alpha}_D(Q^{\ast},\tilde f)=d^{\alpha \beta}K^C_{\beta}G_{DC}, \;\;\;{\mathscr A}^{\alpha}_p(Q^{\ast},\tilde f)=d^{\alpha \beta}K^a_{\beta}G_{ap}.$$

In this case,   ${\tilde G}_{\tilde {\mathcal A} \tilde {\mathcal B}}$ is given by the following expression:
\begin{equation}
\displaystyle
\left(
\begin{array}{ccc}
(P_{\bot})^{A'}_A(P_{\bot})^{B'}_B\bigl(\tilde G^{\rm H}_{A'B'}+d_{\mu \nu }{\mathscr A}_{A'}^\mu {\mathscr A}_{B'}^\nu \bigr) & 0 & (P_{\bot})^{A'}_A{\mathscr A}_{A'}^\mu
d_{\mu\nu}
\bar{u}^{\nu}_{\alpha} \\ 
0 & G_{ab}  & {\mathscr A}^{\mu}_a d_{\mu\nu}{\bar u}^{\nu}_{\alpha}\\
(P_{\bot})^{B'}_B{\mathscr A}^{\mu}_{B'}d_{\mu\nu}{\bar u}^{\nu}_{\beta} & {\mathscr A}^{\mu}_b d_{\mu\nu}{\bar u}^{\nu}_{\beta} & d_{\mu \nu}{\bar u}^{\mu}_{\alpha}{\bar u}^{\nu}_{\beta}\\
\end{array}
\!\!\right)\!.
\label{transfmetric}
\end{equation}

The pseudoinverse matrix  to the matrix (\ref{transfmetric}) is 
\begin{equation}
 \displaystyle
\left(
\begin{array}{ccc}
 G^{A'B'}N^A_{A'}N^B_{B'} & G^{A'B'}N^C_{A'}N^B_{B'}\underset{\scriptscriptstyle{(\gamma)}}{{\mathscr A}^{\mu}_C} K^a_{\mu}  & -G^{A'B'}N^C_{A'}N^B_{B'}\,\underset{\scriptscriptstyle{(\gamma)}}{{\mathscr A}^{\beta}_C} \bar v ^{\alpha}_{\beta} \\
G^{A'B'}N^A_{A'}N^C_{B'}\underset{\scriptscriptstyle{(\gamma)}}{{\mathscr A}^{\mu}_C} K^b_{\mu}  & G^{AB}N^a_AN^b_B+G^{ab} & -G^{EC}{\Lambda}^{\beta}_E{\Lambda}^{\mu}_CK^b_{\mu}\bar v ^{\alpha}_{\beta}
\\
-G^{A'B'}N^A_{A'}N^C_{B'}\underset{\scriptscriptstyle{(\gamma)}}{{\mathscr A}^{\varepsilon}_C}\bar v ^{\beta}_{\varepsilon} & -G^{EC}{\Lambda}^{\varepsilon}_E{\Lambda}^{\mu}_CK^a_{\mu}\bar v ^{\beta}_{\varepsilon} & G^{BC}{\Lambda}^{\alpha'}_B{\Lambda}^{\beta'}_C\bar v ^{\alpha}_{\alpha'}v ^{\beta}_{\beta'}
\end{array}
\!\!\right)\!,
\label{invers_metric}
\end{equation}
where $\underset{\scriptscriptstyle{(\gamma)}}{{\mathscr A}^{\mu}_C}=\gamma^{\mu\nu}K^B_{\nu}G_{BC}$, $N^a_B=-K^a_{\mu}{\Lambda}^{\mu}_B$ is  the component of the projection operator $N^{\tilde A}_{\tilde B}=(N^A_B,0,N^a_B,\delta^a_b)$, (in our notation, the index $\tilde A=(A,a)$). Since $N^{\tilde A}_{\tilde B}K^{\tilde B}_{\alpha}=0$, $N^{\tilde A}_{\tilde B}$ projects onto planes perpendicular to the gauge orbits in ${\rm P}(\tilde {\mathcal M},\mathcal G)$.

Note that the elements  of the matrix (\ref{invers_metric}) can also be rewritten somewhat differently, using the following identities:
$$G^{BB'}{\Lambda}^{\alpha}_B{\Lambda}^{\beta}_{B'}=(\gamma^{\alpha \beta}+G^{A'B'}N^A_{A'}N^B_{B'}\underset{\scriptscriptstyle{(\gamma)}}{{\mathscr A}^{\alpha}_A}\underset{\scriptscriptstyle{(\gamma)}}{{\mathscr A}^{\beta}_B}),$$
$$G^{AB}N^a_AN^b_B+G^{ab}=\bigl(\gamma^{\alpha \beta}+G^{A'B'}N^A_{A'}N^B_{B'}\underset{\scriptscriptstyle{(\gamma)}}{{\mathscr A}^{\alpha}_A}\underset{\scriptscriptstyle{(\gamma)}}{{\mathscr A}^{\beta}_B}\bigr)K^a_{\alpha}K^b_{\beta}+G^{ab}.$$

\section{Stochastic differential equations in  coordinates of the principal fiber bundle}
Transition from the original coordinates $(Q^A, f^a)$, given on the local chart of the manifold  $\tilde {\mathcal P}$, to the special coordinates $(Q^{\ast}{}^A, \tilde f ^b, a^{\alpha})$ leads to the need for a corresponding replacement of the original stochastic process $\eta_t$ generating the measure in the path integral (\ref{orig_path_int}).
The local components $(Q^{\ast}_t{}^A, \tilde f^b_t, a^{\alpha}_t )$ of a new process must be solutions of those local stochastic differential equations that can be obtained from the equations (\ref{eta_1}) and (\ref{eta_2}) as a result of the change of variables performed in stochastic processes. These equations can be derived  using the It\^{o}'s differentiation formula.

Consider, for example, how to obtain a stochastic differential equation for the component $Q^{\ast}{}^A(t) (\equiv Q^{\ast}_t{}^A)$ of a new stochastic process.
 We know that the local coordinate $Q^{\ast}{}^A$ is defined by the following equation $Q^{\ast}{}^A=F^A(Q,a^{-1}(Q ))$ . The same equation must hold for the local stochastic variable $Q^{\ast}_t{}^A$ and the stochastic variable $\eta^A_1(t)$: $\eta_1^A(t)=F^ A ( Q ^ {\ast}{}^A(t),a^{\alpha }(t))$. Then, applying the It\^{o}'s differentiation formula to  the stochastic variable $Q^{\ast A}_t$, we obtain
 \[
 dQ^{\ast}_t{}^A=\Bigl(\frac{\partial Q^{\ast}{}^A}{\partial Q^B}\Bigr)d\eta^B_1(t)+\frac12\frac{\partial^2 Q^{\ast}{}^A}{\partial Q^C\partial Q^D
}<d\eta^C_1(t),d\eta^D_1(t)>.
\]

Using  the stochastic differential equation (\ref{eta_1}) for $\eta_1^A(t)$ in this equation and re-expressing $\eta_1^A(t)$ through $Q^{\ast}{}^A(t)$  and $a^{\alpha}(t)$ 
in all terms depending on $\eta_1^A(t)$ on the right-hand side of the above equation, we obtain  expressions for the drift and  diffusion coefficients of the stochastic differential equation for $Q^{\ast}_t{}^A$. 
A similar approach can be used to write down the stochastic differential equations for the processes $\tilde f^b_t$ and $a^{\alpha}_t$.

Another way to derive stochastic differential equations for transformed variables is based on the use of the Laplace-Beltrami operator which is  obtained from the original differential operator of the equation  (\ref{1}) as a result of introducing new coordinates   on the manifold $\tilde{\mathcal P}$.  
Namely,  the drift coefficients of these equations will be represented by the coefficients at the first derivatives of thus obtained Laplace-Beltrami operator, and the diffusion coefficients can be found from  the coefficients at the second partial derivatives of this operator.   Using this method, we get the following stochastic differential equations:
\begin{eqnarray}
&&dQ^{\ast}_t{}^B=\frac12(\mu^2\kappa)\Bigl[\frac{1}{\sqrt{dH}}\frac{\partial}{\partial Q^{\ast}{}^A}\Bigl(\sqrt{dH}G^{A'B'}N^A_{A'} N^B_{B'}\Bigr)\nonumber\\
&&+\frac{1}{\sqrt{dH}}\frac{\partial}{\partial \tilde f^a}\Bigl(\sqrt{dH}G^{A'B'}N^C_{A'} N^B_{B'}\underset{\scriptscriptstyle{(\gamma)}}{{\mathscr A}^{\mu}_C}K^a_{\mu}\Bigr)\Bigr]dt
+\mu\sqrt{\kappa}N^B_C\mathscr{X}^C_{\bar M}dw^{\bar M}_t,
\label{sde_Q1}
\end{eqnarray}
where $\mathscr{X}^A_{\bar M}(Q^{\ast}_t)$ is determined from the local equality $\sum_{\bar M=1}^{n_{\mathcal P}}\mathscr{X}^A_{\bar M}\mathscr{X}^B_{\bar M}=G^{AB}$.
\begin{eqnarray}
 &&d\tilde f^b_t=\frac12(\mu^2\kappa)\Bigl[\frac{1}{\sqrt{dH}}\frac{\partial}{\partial Q^{\ast}{}^A}\Bigl(\sqrt{dH}G^{A'B'}N^A_{A'} N^B_{B'}\underset{\scriptscriptstyle{(\gamma)}}{{\mathscr A}^{\mu}_B}K^b_{\mu}\Bigr)\nonumber\\
&&\;\;\;\;\;\;\;\;\;+\frac{1}{\sqrt{dH}}\frac{\partial}{\partial \tilde f^a}\Bigl(\sqrt{dH}(G^{ab}+G^{AB}N^a_AN^b_B)\Bigr)\Bigr]dt
\nonumber\\
&&\;\;\;\;\;\;\;\;\;+\mu\sqrt{\kappa}(N^b_C\mathscr{X}^C_{\bar M}dw^{\bar M}_t+\mathscr{X}^b_{\bar c}dw^{\bar c}_t),
\label{sde_f1}
\end{eqnarray}
and where $\mathscr{X}^a_{\bar c}$ is defined by the local equality $\sum_{\bar c=1}^{n_{\scriptscriptstyle {\mathcal V}}}\mathscr{X}^a_{\bar c}\mathscr{X}^b_{\bar c}=
G^{ab}$.

\begin{eqnarray}
 &&da^{\alpha}_t=\frac12(\mu^2\kappa)\Bigl[\frac{1}{\sqrt{dH}}\frac{\partial}{\partial Q^{\ast}{}^A}\Bigl(-\sqrt{dH}\,G^{EF}N^A_EN^C_F\underset{\scriptscriptstyle{(\gamma)}}{{\mathscr A}^{\beta}_C}\Bigr)\bar v^{\alpha}_{\beta}\nonumber\\
&&\;\;\;\;\;\;\;\;\;+\frac{1}{\sqrt{dH}}\frac{\partial}{\partial \tilde f^b}\Bigl(-\sqrt{dH}\,G^{EC}\Lambda ^{\beta}_E\Lambda^{\mu}_CK^b_{\mu}\Bigr)\bar v^{\alpha}_{\beta}\nonumber\\
&&\;\;\;\;\;\;\;\;\;+G^{BC}\Lambda^{\alpha'}_B \Lambda^{\beta'}_C\bar v^{\beta}_{\beta'}\frac{\partial}{\partial a^{\beta}}\bar v^{\alpha}_{\alpha'}\Bigr]dt
+\mu\sqrt{\kappa}\Lambda_C^{\beta}\bar v^{\alpha}_{\beta}\mathscr{X}^C_{\bar M}dw^{\bar M}_t.
\label{sde_a1}
\end{eqnarray}
A short notation of the above stochastic differential equations is as follows:
\begin{align*}
 dQ^{\ast}_t{}^B&=(\mu^2\kappa)b^B_tdt+\mu\sqrt{\kappa} N^B_C\mathscr{X}^C_{\bar M}dw^{\bar M}_t,
\nonumber\\
 d\tilde f^a_t&=(\mu^2\kappa)b^a_tdt+\mu\sqrt{\kappa}(N^a_C\mathscr{X}^C_{\bar M}dw^{\bar M}_t+\mathscr{X}^a_{\bar c}dw^{\bar c}_t),
\nonumber\\
da^{\alpha}_t&=(\mu^2\kappa)a^{\alpha}_tdt+\mu\sqrt{\kappa}\Lambda_C^{\beta}\bar v^{\alpha}_{\beta}\mathscr{X}^C_{\bar M}dw^{\bar M}_t.
\end{align*}
The drift coefficients of the first two equations can be rewritten to obtain the  stochastic differential equations, which  are used in stochastic theory to describe diffusion on a manifold embedded in some ambient manifold (cf. ref.\cite{Lewis}).

It can be shown that these drift coefficients are equal to the coefficients standing at the first partial derivatives with respect to $Q^{\ast}{}^A$ and $\tilde f^b$ in the Laplacian obtained by regrouping the terms of the original Laplace-Beltrami operator  in the following form:\footnote{Here and in what follows, we adopt the notation in which
subscripts (and superscripts), indicated by capital letters with a tilde, denote multi-indices consisting of two indices, for example  $\tilde B=(B,b)$. It is also assumed that summation occurs over these repeated indices.} 
\begin{eqnarray}
 \lefteqn{\frac12{\tilde G}^{\tilde C\tilde B} \nabla_{\tilde C}\nabla_{\tilde B}= } \nonumber\\
 & &\frac12\Bigl(\tilde{\Pi}^{\tilde A}_{\tilde C}\,{\tilde G}^{\tilde C\tilde B} \nabla_{\tilde A}\nabla_{\tilde B}-K^{\tilde A}_{\alpha}d^{\alpha\beta}(\nabla_{\tilde A}K^{\tilde B}_{\beta})\nabla_{\tilde B}+K^{\tilde A}_{\alpha}d^{\alpha\beta}\nabla_{\tilde A}K^{\tilde B}_{\beta}\nabla_{\tilde B}\Bigr),
\label{laplace_for_j12}
\end{eqnarray}
 where it is also necessary to replace the local coordinates  $(Q^A,f^b)$ with the coordinates $(Q^{\ast}{}^A,\tilde f^b,a^{\alpha})$. 
 
 In the first term on the right side of this equation, the horizontal projection operator 
 \[
 \tilde \Pi=\left(\begin{array}{cc}
        \tilde{\Pi}^D_C &  \tilde{\Pi}^D_a\\
 \tilde{\Pi}^b_C &  \tilde{\Pi}^b_a
       \end{array}\right)
\]
such that $\tilde{\Pi}^{\tilde D}_{\tilde C}K^{\tilde C}_{\alpha}=0$, with  the components
\begin{align*}
\tilde{\Pi}^D_C&=\delta ^D_C-K^D_{\alpha}d^{\alpha \beta}K^B_{\beta}G_{BC},\;\;\tilde{\Pi}^D_a=-K^D_{\alpha}d^{\alpha \beta}K^b_{\beta}G_{ba},\\
\tilde{\Pi}^b_C&=-K^b_{\alpha}d^{\alpha \beta}K^B_{\beta}G_{BC},\;\;\;\;\;\;\;\;\,\tilde{\Pi}^b_a=\delta^b_a-K^b_{\alpha}d^{\alpha \beta}K^q_{\beta}G_{qa},
\end{align*}
 has the following properties:
\[
 \tilde{\Pi}^{\tilde A}_{\tilde L}N^{\tilde L}_C=\tilde{\Pi}^{\tilde A}_C, \;\;\tilde{\Pi}^{E}_{B}N^{R}_E=N^R_B,\;\;\tilde{\Pi}^{E}_{a}N^{R}_E=0.
\]

 Note also that  in our case
\[
 \bigl({\tilde G}^{\tilde C\tilde B} \nabla_{\tilde C}\nabla_{\tilde B}\bigr)(Q,f)=\bigl(G^{AB}\nabla_A\nabla_B\bigr)(Q)+G^{pq}\bigl(\nabla_p\nabla_q\bigr)(f).
\]
The symbol $\nabla_{A}$ denotes the covariant derivative defined using the Christoffel symbols of the manifold $\mathcal P$. However, due to the special choice of the initial metric on the manifold $\tilde {\mathcal P}$, in our case we have $\nabla_q(f)=\frac{\partial}{\partial f^q}$.

As a result of the calculations, it follows that in each of the equations (\ref{sde_Q1}) and (\ref{sde_f1}), the drift coefficients  are represented by the sum of two terms, $b_{\Roman 1}$ and $b_{\Roman 2}$, coming only from the first and the second terms on the right-hand side  of the equation (\ref{laplace_for_j12}). The third term  does not contribute to the drift coefficients of stochastic differential equations on the $\tilde \Sigma $.

\subsection{The drift coefficient $b_{\Roman 2}$ as a projection of  the mean curvature vector field of the orbit  onto the submanifold $\tilde \Sigma$}

In this subsection we consider the geometry of the drift term $b_{\Roman 2}$. Our calculation show that this term is given by the projection on the submanifold  $\tilde\Sigma $ of the mean curvature vector of the orbit.

In our case, the Killing vector fields
which in the coordinates $(Q^A,f^a)$ are given by the formula
$$\tilde K_{\alpha}=
     K^{A}_{\alpha }(Q)\frac{\partial}{\partial Q^{A}}+K^{a}_{\alpha }(f)\frac{\partial }{\partial f^a},$$ are tangent  to the group orbit.

The mean curvature vector field (mean curvature normal) of the orbit is defined as follows:
\begin{eqnarray}
 &&\vec{\mu}=\frac12d^{\alpha\beta}\biggl[\Bigl(\tilde{\Pi}^D_C(\nabla_{K_{\alpha}}K_{\beta})^C+\tilde{\Pi}^D_a(\nabla_{K_{\alpha}}K_{\beta})^a\Bigr)
\frac{\partial}{\partial Q{}^{ D}}
\nonumber\\
&&\;\;\;\;\;\;\;+\Bigl(\tilde{\Pi}^b_C(\nabla_{K_{\alpha}}K_{\beta})^C+\tilde{\Pi}^b_a(\nabla_{K_{\alpha}}K_{\beta})^a\Bigr)\frac{\partial}{\partial f^b}\biggr].
\label{mean_curv_Qf}
\end{eqnarray}
By $(\nabla_{K_{\alpha}}K_{\beta})^C(Q)$, used in $\vec{\mu}$, we denote
\[
 K^A_{\alpha}(Q)\frac{\partial}{\partial Q^A}K^C_{\beta}(Q)+K^A_{\alpha}(Q)K^B_{\beta}(Q)\Gamma^C_{AB}(Q),
\]
where
\[
 \Gamma^C_{AB}(Q)=\frac12 G^{CE}(Q)\Bigl(\frac{\partial}{\partial Q^A}G_{EB}(Q)+\frac{\partial}{\partial Q^B}G_{EA}(Q)-\frac{\partial}{\partial Q^E}G_{AB}(Q)\Bigr).
\]

To obtain the projection of $\vec{\mu}$ onto the submanifold ${\tilde \Sigma}$, one must first find how the terms on the right-hand side of the equation (\ref{mean_curv_Qf} ) are expressed in terms of the coordinates $(Q{} ^{ \ast A},\tilde f^a, a^{\alpha})$. It can be shown that
$\bigl(\nabla_{K_{\alpha}}K_{\beta}\bigr)^b(f)$ is transformed as 
\[
 \bigl(\nabla_{K_{\alpha}}K_{\beta}\bigr)^b(f)={\rho}^{\mu}_{\alpha}(a){\rho}^{\nu}_{\beta}(a)K^q_{\mu}(\tilde f)(\bar J_{\nu})^c_q\bar D^b_c(a).
\]
In derivation of this representation it was used the  formula
\[
 D^q_b(a)(\bar J_{\nu})^b_p\bar D^p_c(a)={\rho}^{\beta}_{\nu}(a)(\bar J_{\beta})^q_c.
\]
 For $\bigl(\nabla_{K_{\alpha}}K_{\beta}\bigr)^{A}(Q)$ we have the  following expression:
\[
 \bigl(\nabla_{K_{\alpha}}K_{\beta}\bigr)^A(Q)=F^A_DK^D_{\mu}(Q^{\ast})c^{\nu}_{\alpha \beta}\rho ^{\mu}_{\nu}+F^A_D\rho ^{\sigma}_{\alpha}\rho ^{\mu}_{\beta}(\nabla_{K_{\sigma}}K_{\mu}\bigr)^D(Q^{\ast}).
\]
 As a result of the performed transformations, the mean curvature vector field $\vec{\mu}(Q{}^{\ast},\tilde f, a)$ is represented as 
\begin{eqnarray}
 &&\vec{\mu}=\frac12d^{\alpha \beta}\bigl(\nabla_{K_{\alpha}}K_{\beta}\bigr)^B(Q^{\ast})\Bigl(N^R_B\frac{\partial}{\partial Q{}^{\ast R}}+{\Pi}^C_B\Lambda ^{\mu}_C\bar v^{\nu}_{\mu}(a)\frac{\partial}{\partial a^{\nu}}+N^b_B\frac{\partial}{\partial \tilde f^b}\Bigr)
\nonumber\\
&&\;\;\;\;+\frac12d^{\alpha \beta}\bigl(\nabla_{K_{\alpha}}K_{\beta}\bigr)^b(\tilde f)\Bigl({\Pi}^E_b\Lambda ^{\mu}_E\bar v^{\nu}_{\mu}(a)\frac{\partial}{\partial a^{\nu}}+\frac{\partial}{\partial \tilde f^b}\Bigr).
\label{mean_curv}
 \end{eqnarray}
(We recall that $N^b_B=-\Lambda ^{\mu}_BK^b_{\mu}.$)

The projection  of  the vector field $\vec{\mu}$ $(Q{}^{\ast},\tilde f, a)$ on the submanifold   $\tilde\Sigma \in \mathcal P\times \mathcal V$  is defined by the following formula: 
\begin{eqnarray*}
 &&{\tilde G}^{SL}{\tilde G}\Bigl(\vec{\mu},\frac{\partial}{\partial Q{}^{\ast S}}\Bigr)\frac{\partial}{\partial Q{}^{\ast L}}+
{\tilde G}^{Sq}{\tilde G}\Bigl(\vec{\mu},\frac{\partial}{\partial Q{}^{\ast S}}\Bigr)\frac{\partial}{\partial {\tilde f}^q}
\nonumber\\
&&+{\tilde G}^{aq}{\tilde G}\Bigl(\vec{\mu},\frac{\partial}{\partial {\tilde f}^a}\Bigr)\frac{\partial}{\partial {\tilde f}^q}
+{\tilde G}^{aL}{\tilde G}\Bigl(\vec{\mu},\frac{\partial}{\partial {\tilde f}^a}\Bigr)\frac{\partial}{\partial Q{}^{\ast L}}.
\end{eqnarray*}
Applying this formula, we get the following representation for $\vec b_{\Roman 2}(Q^{\ast},\tilde f)$:
\begin{eqnarray}
&&\vec b_{\Roman 2}=-\frac12 \,G^{CC'}N^L_{C'}N^{B'}_CG_{BB'}\,d^{\alpha \beta}\bigl(\nabla_{K_{\alpha}}K_{\beta}\bigr)^B\frac{\partial}{\partial Q{}^{\ast L}}
\nonumber\\
&&\;\;\;\;\;\;\;\;\;-\frac12d^{\alpha \beta}\Bigl(N^q_B\,\bigl(\nabla_{K_{\alpha}}K_{\beta}\bigr)^B+\bigl(\nabla_{K_{\alpha}}K_{\beta}\bigr)^q\Bigr)\frac{\partial}{\partial {\tilde f}^q}.
\label{b_2}
\end{eqnarray}
Since  the first line of (\ref{b_2}) can be also rewritten as 
\[
 -\frac12 \,d^{\alpha \beta}N^L_B(Q{}^{\ast})\bigl(\nabla_{K_{\alpha}}K_{\beta}\bigr)^B\, \frac{\partial}{\partial Q{}^{\ast L}},
\]
then, using the shorthand notation, $\vec b_{\Roman 2}$ can be written  in the following form:
\[
 \vec b_{\Roman 2}=-\frac12 \,d^{\alpha \beta}N^{\tilde L}_{\tilde B}\bigl(\nabla_{K_{\alpha}}K_{\beta}\bigr)^{\tilde B}\, \frac{\partial}{\partial \tilde Q{}^{{\tilde L}}},
\]
(that is, here $\tilde L=(L,a)$ and  $\tilde Q^{\tilde L}\equiv(Q{}^{{\ast}  L},\tilde f^a)$).

\subsection{The drift coefficient  {$b_{\Roman 1}^A$}}
 In this subsection, we show that the drift coefficient $b_{\Roman 1}^A$ can  be obtained using the corresponding stochastic differential equation given on the orbit space  $\tilde{\mathcal M}$ (which is locally isomorphic to $\tilde{ \Sigma}$). The differential generator of the semigroup associated with the stochastic process described by this stochastic differential equation arises from the first term on the right-hand side of the equation (\ref{laplace_for_j12}) after its  restriction  to the manifold  $\tilde{\mathcal M}$. The differential generator thus obtained is the Laplace-Beltrami operator on $\tilde{\mathcal M}$. 
 
 Note that in our case, the local coordinates on the orbit space manifold $\tilde{\mathcal M}$ are given by the invariant coordinates $(x^i,\tilde f^a)$, where $x^i,i= 1, ... . ,n_{\mathcal M},$ are coordinates on $\mathcal M$. They are defined under the condition that the local submanifold $\Sigma$ can also be represented in the parametric form: $Q^A=Q^{\ast A}(x^i) $.   This implies that the coordinates of points on $\Sigma$ must satisfy the equation $\chi^{\alpha}(Q^{ \ast }{}^A(x^i))=0$.

   The metric tensor of the Riemannian manifold  $\tilde{\mathcal M}$ has the following components in the coordinate basis  $(\frac{\partial}{\partial x^i},\frac{\partial}{\partial \tilde f^a})$:
  \begin{equation}
  \left(
\begin{array}{cc}
\tilde h_{ij} & \tilde G^{\rm H}_{B b}Q_{i}^{*B}\\
\tilde G^{\rm H}_{Aa}Q_{j}^{*A} & \tilde G^{\rm H}_{ba}\\
\end{array}
\right),
\label{metric_h_ija}
\end{equation}
where $\tilde h_{ij}=Q^{\ast}{}^A_i\tilde G{}^{\rm H}_{AB}(Q^{\ast}(x))Q^{\ast}{}^B_j$ with $Q^{\ast}{} ^B_i\equiv\frac{\partial}{\partial x^i}Q^{\ast}{}^B(x)$.
This means that the Riemannian manifold $\tilde{\mathcal M}$ can be considered as a submanifold in the (Riemannian) manifold $(\tilde {\mathcal P},\tilde G^{\rm H}_{\tilde A\tilde B})$ with degenerate metric $\tilde G^{\rm H}_{\tilde A\tilde B}$.

The elements of the inverse matrix to  matrix $(\ref{metric_h_ija})$ are given by
\begin{align*}
 &\tilde h^{ij}=h^{ij}:=h^{SD}T^i_DT^j_S=G^{EF}N^S_EN^D_FT^i_DT^j_S,\;\;\tilde h^{jb}:=h^{bP}T^j_P=G^{EF}N^b_FN^P_ET^j_P,\\
 &\tilde h^{ab}:=h^{ab}=G^{ab}+G^{EF}N^a_EN^b_F,
 \end{align*}
where the  operator
$$T^i_D=(P_{\bot})^B_D(Q^{\ast}(x))G^{\rm H}_{BL}(Q^{\ast}(x))Q^{\ast L}_m(x) h^{mi}(x),$$ with the properties $T^i_DQ^{\ast D}_j=\delta^i_j$ and $T^i_DQ^{\ast B}_i=(P_{\bot})^B_D$, is defined by using the projection operator $P_{\bot}$ on the tangent plane to the submanifold $\Sigma$ of the manifold $\mathcal P$.

The diffusion on $\tilde{\mathcal M}$ is given by the stochastic process locally represented by the processes $(x^i_t,\tilde f^a_t)$. 
But our goal is to obtain a description of diffusion on $\tilde{\mathcal M}$ in terms of the local stochastic differential equations for stochastic variables associated with the ambient manifold, that is in our case for the 
stochastic variable $Q_t^{\ast}{}^A$, which is associated with to the dependent variable $Q^{\ast}{}^A$, and for $\tilde f^a_t$. 

Since  on $\;\tilde{\mathcal M}$ the variable $Q^{\ast}{}^A$ is now the function of $x^i$, the stochastic differential equation for $Q^{\ast}{}^A(x(t))$ can be obtained by means of  the replacement of the variables  in the stochastic differential equations: 
\begin{equation}
 dQ^{\ast}{}^A(x(t))=Q^{\ast}{}^A_i(x(t)) dx^i_t +\frac12 Q^{\ast}{}^A_{ij}(x(t))<dx^i_t,dx^j_t>.
\label{sde_Q_ast(x)}
\end{equation}
In turn, the drift coefficient $b_{\Roman{1}}^i(t)$ of the stochastic differential equation for $x_t$ 
\[
 dx^i_t=\mu^2\kappa\,b_{\Roman{1}}^i(t)dt+\mu\sqrt{\kappa}X^i_{\bar{n}}(t)dw^{\bar{n}}_t
\]
can be easily found using the Laplace-Beltrami operator of the manifold  $\tilde{\mathcal M}$:
\[
 b_{\Roman{1}}^i=\frac12\Bigl(\frac{1}{H^{1/2}}\frac{\partial }{\partial x^j}\bigl(H^{1/2}h^{ij}\bigr)+\frac{1}{H^{1/2}}\frac{\partial }{\partial \tilde f^a}\bigl(H^{1/2}\tilde h^{ai}\bigr)\Bigr).
\]
On the other hand, 
\begin{equation}
 b_{\Roman{1}}^i=-\frac12\bigl(h^{kn}{\Gamma}^i_{kn}+\tilde h^{kb}{\Gamma}^i_{kb}+\tilde h^{am}{\Gamma}^i_{am}+\tilde h^{ab}{\Gamma}^i_{ab}\bigr),
\label{b_iRoman_1}
\end{equation}
where ${\Gamma}$ are   Christoffel symbols of  $\tilde{\mathcal M}$ with the metric tensor (\ref{metric_h_ija}).

The Christoffel symbol ${\Gamma}^i_{kn}$ is defined as
\begin{equation}
 {\Gamma}^i_{kn}=\tilde h^{im}{\Gamma}_{knm}+\tilde h^{ia}{\Gamma}_{kna},
\label{Chr_gikn}
\end{equation}
where
\begin{eqnarray*}
 &&{\Gamma}_{knm}=\frac12(\tilde h_{km,n}+\tilde h_{nm,k}-\tilde h_{kn,m}),
\nonumber\\
&& {\Gamma}_{kna}=\frac12(\tilde h_{ka,n}+\tilde h_{na,k}-\tilde h_{kn,a}).
\nonumber\\
\end{eqnarray*}
Note that $\tilde h_{kn,a}$ in the previous formula is the partial derivative of $\tilde h_{kn}$ with respect to $\tilde f^a$.
  Using explicit expressions for the  components of the metric given on the orbit space manifold, one can obtain the following representations for the above Christoffel symbols:
\begin{eqnarray}
 &&{\Gamma}_{knm}={}^{ \mathrm  H}{\tilde \Gamma}_{BMT}Q^{\ast}{}^B_kQ^{\ast}{}^M_nQ^{\ast}{}^T_m+{\tilde G}^H_{AB}Q^{\ast}{}^A_{kn}Q^{\ast}{}^B_m,
\nonumber\\
&&{\Gamma}_{kna}={}^{ \mathrm  H}{\tilde \Gamma}_{BMa}Q^{\ast}{}^B_kQ^{\ast}{}^M_n+{\tilde G}^H_{Aa}Q^{\ast}{}^A_{mn},
\label{Chr_GHkn(ma)}
\end{eqnarray}
where
\[
{}^{ \mathrm  H}{\tilde \Gamma}_{BMD}\equiv\frac12({\tilde G}^{\rm H}_{BD,M}+{\tilde G}^{\rm H}_{MD,B}-{\tilde G}^{\rm H}_{BM,D}),
\]
\[
{}^{ \mathrm  H}{\tilde \Gamma}_{BMa}\equiv\frac12({\tilde G}^{\rm H}_{Ba,M}+{\tilde G}^{\rm H}_{MD,B}-{\tilde G}^{\rm H}_{BM,a}).
\]
To determine the
 Christoffel symbol 
 ${}^{ \mathrm  H}{\tilde \Gamma}^{\tilde R}_{BM}$ we use the following equality: 
\begin{equation}
 {}^{ \mathrm  H}{\tilde \Gamma}_{BM\tilde D}={\tilde G}^{\mathrm H}_{\tilde R\tilde D}{}^{ \mathrm  H}{\tilde \Gamma}^{\tilde R}_{BM}.
\label{Chr_H_BMD}
\end{equation}
Note, however, that this defines ${}^{ \mathrm  H}{\tilde \Gamma}^{\tilde R}_{BM}$ only modulo such terms
 $T^{\tilde M}_{BC}$ that satisfy  ${\tilde G}^{\mathrm H}_{A\tilde M}T^{\tilde M}_{BC}=0$.
In addition, note  that by our notation (\ref{Chr_H_BMD}) also means that 
\begin{equation}
 {}^{ \mathrm  H}{\tilde \Gamma}_{BMa}={\tilde G}^{\mathrm H}_{\tilde Ra}{}^{ \mathrm  H}{\tilde \Gamma}^{\tilde R}_{BM}.
\label{Chr_H_BMa}
\end{equation}
Substituting  (\ref{Chr_H_BMD}) and (\ref{Chr_H_BMa}) into (\ref{Chr_GHkn(ma)}), and then the result  in (\ref{Chr_gikn}),  we obtain, after appropriate  transformations, the following representation for ${\Gamma}^i_{kn}$:
\[
 {\Gamma}^i_{kn}=T^i_SN^S_R \bigl({}^{ \mathrm  H}{\tilde \Gamma}^{ R}_{BM}Q^{\ast}{}^B_kQ^{\ast}{}^M_n+Q^{\ast}{}^R_{kn}\bigr).
\]
Such a representation is obtained using the following identities:
\[
 N^{\tilde D}_F{\tilde G}^{\rm H}_{\tilde R\tilde D}={\tilde G}^{\rm H}_{\tilde R F},\;\;\;
G^{EF}{\tilde G}^{\rm H}_{\tilde R F}={\tilde \Pi}^E_{\tilde R},\;\;\;
N^A_E{\tilde \Pi}^E_{\tilde R}=N^A_R,\;\;\;N^A_E{\tilde \Pi}^E_{r}=0.
\]
In the same way one  can get that 
\begin{eqnarray*}
 &&{\Gamma}^i_{kb}=T^i_SN^S_R \,{}^{ \mathrm  H}{\tilde \Gamma}^{ R}_{Ab}Q^{\ast}{}^A_k,\;\;{\Gamma}^i_{am}=T^i_SN^S_R \,{}^{ \mathrm  H}{\tilde \Gamma}^{ R}_{aB}Q^{\ast}{}^B_m,\;\;
{\Gamma}^i_{ab}=T^i_SN^S_R \,{}^{ \mathrm  H}{\tilde \Gamma}^{ R}_{ab}.
\nonumber\\
\end{eqnarray*}
Using the obtained Christoffel symbols in (\ref{b_iRoman_1}), it can be shown that the expression for the drift coefficient $b_{\Roman{1}}^A$ of the equation  (\ref{sde_Q_ast(x)}) is given by the following formula:
\begin{eqnarray} 
 &&b_{\Roman{1}}^A=-\frac12N^A_R\Bigl(G^{EF}N^B_EN^M_F\,{}^{ \mathrm  H}{\tilde \Gamma}^{ R}_{BM}+G^{EF}N^b_FN^B_E\,{}^{ \mathrm  H}{\tilde \Gamma}^{ R}_{Bb}+G^{EF}N^a_EN^B_F\,{}^{ \mathrm  H}{\tilde \Gamma}^{ R}_{aB}
 \nonumber\\
 &&\;\;\;\;\;\;\;\;+(G^{ab}+G^{EF}N^a_EN^b_F)\,{}^{ \mathrm  H}{\tilde \Gamma}^{ R}_{ab}+ h^{kn}Q^{\ast}{}^R_{kn}\Bigr)+\frac12 h^{ij}Q^{\ast}{}^A_{ij},
\label{b_1_A_Q}
 \end{eqnarray}
where the  terms   depend on $Q^{\ast}(x)$ and $\tilde f$ and, therefore, in (\ref{sde_Q_ast(x)}) on $Q^{\ast}(x_t)$ and $\tilde f_t$.
Also note that,  using a shorthand notation, $b_{\Roman{1}}^A$ can be rewritten  as 
\begin{equation}
 b_{\Roman{1}}^A=-\frac12N^A_R\Bigl(h^{\tilde B \tilde M}\,{}^{ \mathrm  H}{\tilde \Gamma}^{ R}_{\tilde B\tilde M}+ h^{kn}Q^{\ast}{}^R_{kn}\Bigr)+\frac12 h^{ij}Q^{\ast}{}^A_{ij}.
\label{b_1_A_Q_short}
\end{equation}

\subsection{The drift coefficient $b_{\Roman 1}^a$}
The drift coefficient $b_{\Roman 1}^a$ of the stochastic differential equation for the stochastic variable $\tilde f^a_t$ can also be  found using the Laplace-Beltrami operator of the manifold $\tilde{\mathcal M}$. 
 This term is given by the following expression:
\begin{eqnarray*}
&&b^a_{\Roman 1}=\frac12\Bigl(\frac{1}{H^{1/2}}\frac{\partial }{\partial x^i}\bigl(H^{1/2}\tilde h^{ia}\bigr)+\frac{1}{H^{1/2}}\frac{\partial }{\partial \tilde f^b}\bigl(H^{1/2}\tilde h^{ab}\bigr)\Bigr).
\nonumber\\
\end{eqnarray*}
This expression can be rewritten using   Christoffel symbolls as follows:
\[
 b^a_{\Roman 1}=-\frac12\bigl(h^{kn}{\Gamma}^a_{kn}+\tilde h^{kb}{\Gamma}^a_{kb}+\tilde h^{bm}{\Gamma}^a_{bm}+\tilde h^{bc}{\Gamma}^a_{bc}\bigr).
\]
We omit here the steps with the necessary transformations to obtain the representation of $b^a_{\Roman 1}$, since they are similar to what we did when obtaining the drift coefficient $b^A_{\Roman 1}$.  The relationships between Christoffel symbols required for these transformations are given for reference in  Appendix.

As a result, we  get the following representation for $b^a_{\Roman 1}$: 
\begin{eqnarray}
 &&b^a_{\Roman 1}=-\frac12N^a_R\bigl(G^{EF}N^{ B}_EN^{ M}_F\,{}^{\mathrm H}{\tilde \Gamma}^R_{ B M}
 +G^{EF}N^{ B}_EN^{b}_F\,{}^{\mathrm H}{\tilde \Gamma}^R_{ bB}
 +G^{EF}N^{ B}_EN^{ b}_F\,{}^{\mathrm H}{\tilde \Gamma}^R_{ B b}
 \nonumber\\
 &&\;\;\;\;\;\;\;\;+G^{EF}N^{b}_EN^{ c}_F\,{}^{\mathrm H}{\tilde \Gamma}^R_{ bc}+G^{bc}\,{}^{\mathrm H}{\tilde \Gamma}^R_{ bc}
 +h^{kn}Q^{\ast}{}^R_{kn}\bigr)
 \nonumber\\
 &&\;\;\;\;\;\;\;\;-\frac12\bigl(G^{EF}N^{\tilde B}_EN^{ \tilde M}_F\,{}^{\mathrm H}{\tilde \Gamma}^a_{\tilde B \tilde M}
 +G^{bc}\,{}^{\mathrm H}{\tilde \Gamma}^a_{ bc}\bigr).
 \label{b_1_a_Q}
 \end{eqnarray}
The same expression can be rewritten as
\[
 b^a_{\Roman 1}=-\frac12N^a_{\tilde R}h^{\tilde B \tilde M}\,{}^{\mathrm H}{\tilde \Gamma}^{\tilde R}_{\tilde B\tilde M}-\frac12N^a_A h^{kn}Q^{\ast}{}^A_{kn}.
\]
Note that when getting (\ref{b_1_a_Q}), the following identities were used:
$$N^T_F{\tilde G}^{\mathrm H}_{\tilde RT}+N^b_F{\tilde G}^{\mathrm H}_{\tilde Rb}={\tilde G}^{\mathrm H}_{\tilde RF},$$
\[
 G^{ab}{\tilde G}^{\mathrm H}_{Ab}=\tilde{\Pi}^a_A,\;\;\;G^{EF}{\tilde G}^{\mathrm H}_{AF}=\tilde{\Pi}^E_A,\;\;\;N^a_E\tilde{\Pi}^E_A+\tilde{\Pi}^a_A=N^a_A.
\]

From (\ref{b_1_A_Q}) and (\ref{b_1_a_Q}) we see that the expressions for $b^A_{\Roman 1}$ and $b^a_{\Roman 1}$ include still ``untransformed'' terms with second partial derivatives of $Q^{\ast}{}^A(x)$ with respect to $x^k$ and $x^n$. It turns out that the solution to this problem can be found using the mean curvature  normal  of the orbit space $\tilde{\mathcal M}$.

The mean curvature normal is defined as the trace of the second fundamental form. In our case, the local basis (the frame)  on the tangent plane to the manifold  of the orbit space  (which is the submanifold in the ambient space with the degenerated metric)
is given by the coordinate vector fields 
$(\frac{\partial}{\partial x^i},\frac{\partial}{\partial \tilde f^a})\equiv(e_i,e_a)$. To calculate  the mean curvature normal we  use the following representation for these coordinate vector fields:
$(Q^{\ast}{}^A_i(x)\frac{\partial}{\partial Q^{\ast}{}^A},\frac{\partial}{\partial \tilde f^a})=(Q^{\ast}{}^A_ie_A,e_a)$,
so the mean curvature normal $\vec{j_{\Roman 1}}$ must be given as 
\[
 \vec{j_{\Roman 1}}=j^A_{\Roman 1}\frac{\partial}{\partial Q^{\ast}{}^A}+j^a_{\Roman 1}\frac{\partial}{\partial \tilde f^a}.
\]

The components of the mean curvature normal $\vec{j_{\Roman 1}}$ are defined as follows:
\begin{equation}
 j^{\tilde A}_{\Roman 1}=\frac12\bigl(\delta^{\tilde A}_{\tilde B}-N^{\tilde A}_{\tilde B}\bigr)\bigl(\dots\bigr)^{\tilde B},
\label{j_1_def}
\end{equation}
where
\[
 \bigl(\dots\bigr)^{\tilde B}=\Bigl(h^{ij}[\nabla_{e_i}e_{j}]+\tilde h^{ia}[\nabla_{e_i}e_{a}]+\tilde h^{ai}[\nabla_{e_a}e_{i}]+\tilde h^{ab}[\nabla_{e_a}e_{b}]\Bigr)^{\tilde B}.
\]

To calculate the mean curvature normal $\vec{j_{\Roman 1}}$, we use the formula (\ref{j_1_def}), which states that $\vec{j_{\Roman 1}}$  is defined as the projection  of the vector field, which in our case is represented by the above expression given in parentheses (...), on the normal bundle to the orbit space\cite{Betounes}. This projection is performed by an operator whose components are specified using $(\delta^{\tilde A}_{\tilde B}-N^{\tilde A}_{\tilde B})$, where $N^{\tilde A}_{\tilde B}=(N^A_B,N^a_B,0,N^a_b=\delta ^a_b)$.

Note that from (\ref{j_1_def}) it follows that
\[
 j^{ A}_{\Roman 1}=\frac12\bigl(\delta^{ A}_{ B}-N^{A}_{B}\bigr)\bigl(\dots\bigr)^{B}\;\;\;{\rm and}\;\;\;\;j^{ a}_{\Roman 1}=-\frac12N^{a}_{B}\bigl(\dots\bigr)^{B}.
\]
In our local basis given by the tangent vector fields, we have 
\begin{eqnarray*}
 \nabla_{e_i}e_{j}&=&\nabla_{Q^{\ast}{}^A_i(x)e_A}Q^{\ast}{}^B_j(x)e_B
 \nonumber\\
 &=&h^{ij}\bigl(Q^{\ast}{}^A_iQ^{\ast}{}^B_j
 \,{}^{\rm H}{\tilde \Gamma}^C_{AB}+Q^{\ast}{}^C_{ij}\bigr)\frac{\partial}{\partial Q^{\ast}{}^C}
 +h^{ij}Q^{\ast}{}^A_iQ^{\ast}{}^B_j
 \,{}^{\rm H}{\tilde \Gamma}^b_{AB}\frac{\partial}{\partial \tilde f^b}
\nonumber\\
&=&\bigl(G^{EF}N^A_EN^B_F
 \,{}^{\rm H}{\tilde \Gamma}^C_{AB}+h^{ij}Q^{\ast}{}^C_{ij}\bigr)\frac{\partial}{\partial Q^{\ast}{}^C}
 +G^{EF}N^A_EN^B_F
 \,{}^{\rm H}{\tilde \Gamma}^b_{AB}\frac{\partial}{\partial \tilde f^b},
 \end{eqnarray*}
where $G^{EF}, N^A_E$, $N^B_F$ depend on $Q^{\ast}(x)$, ${}^{\rm H}{\tilde \Gamma}$ depend on $Q^{\ast}(x)$ and $\tilde f$.
The other components of $\bigl(\dots\bigr)^{\tilde B}$ are found similarly. 

As a result, we get that
\begin{align}
 j^{ a}_{\Roman 1}&=-\frac12N^{a}_{C}\Bigl(G^{EF}N^A_EN^B_F
 \,{}^{\rm H}{\tilde \Gamma}^C_{AB}+G^{EF}N^B_EN^b_F
 \,{}^{\rm H}{\tilde \Gamma}^C_{Bb}+G^{EF}N^b_EN^B_F
 \,{}^{\rm H}{\tilde \Gamma}^C_{bB}
 \nonumber\\
 &\;\;\;\;\;\;\;\;\;+(G^{db}+G^{EF}N^d_EN^b_F)\,{}^{\rm H}{\tilde \Gamma}^C_{db}+h^{ij}Q^{\ast}{}^C_{ij}\Bigr)
 \nonumber\\
 &=-\frac12N^{a}_{C}\bigl(h^{\tilde E \tilde F}\,{}^{\rm H}{\tilde \Gamma}^C_{\tilde E \tilde F}+h^{ij}Q^{\ast}{}^C_{ij}\bigr)
 \label{j_1_a}
 \end{align}
 and
 \begin{align}
 j^{ A}_{\Roman 1}&=\frac12 \bigl(\delta^{ A}_{C}-N^{A}_{C}\bigr)\Bigl(G^{EF}N^{A'}_EN^{B'}_F
 \,{}^{\rm H}{\tilde \Gamma}^C_{A'B'}+G^{EF}N^B_EN^b_F
 \,{}^{\rm H}{\tilde \Gamma}^C_{Bb}
 \nonumber\\
 &\;\;\;\;\;\;\;\;\;+G^{EF}N^b_EN^B_F
 \,{}^{\rm H}{\tilde \Gamma}^C_{bB}+(G^{db}+G^{EF}N^d_EN^b_F)\,{}^{\rm H}{\tilde \Gamma}^C_{db}+h^{ij}Q^{\ast}{}^C_{ij}\Bigr)
 \nonumber\\
 &=\frac12 \bigl(\delta^{ A}_{C}-N^{A}_{C}\bigr)\bigl(h^{\tilde E \tilde F}\,{}^{\rm H}{\tilde \Gamma}^C_{\tilde E \tilde F}+h^{ij}Q^{\ast}{}^C_{ij}\bigr).
 \label{j_1_A}
 \end{align}
Comparing (\ref{b_1_A_Q}) and (\ref{j_1_A}), we conclude that
\begin{equation}
 b^A_{\Roman 1}=-\frac12h^{\tilde B \tilde M}\,{}^{ \mathrm  H}{\tilde \Gamma}^{ A}_{\tilde B\tilde M}+j^{ A}_{\Roman 1}.
\label{b_1_A_Q_new}
\end{equation}
Similarly, from (\ref{b_1_a_Q}) and (\ref{j_1_a}) it follows that
\begin{equation}
 b^a_{\Roman 1}=-\frac12h^{\tilde B \tilde M}\,{}^{ \mathrm  H}{\tilde \Gamma}^{ a}_{\tilde B\tilde M}+j^{ a}_{\Roman 1}.
\label{b_1_a_Q_new}
\end{equation}
Since the mean curvature normal can be also defined by the Weingarten map, i.e.  without using explicit 
coordinate expressions, $\vec{j}_{\Roman 1}$ is the function given on a submanifold $\tilde{\mathcal M}$, so $j^{ A}_{\Roman 1}\equiv j^{ A}_{\Roman 1}(Q^{\ast}(x),\tilde f)$ and $j^{ a}_{\Roman 1}\equiv j^{ a}_{\Roman 1}(Q^{\ast}(x),\tilde f)$.

In our case, this can be shown by using in (\ref{j_1_A}) the identity 
\[ 
N^A_{B,D}\,Q^{\ast}{}^B_iQ^{\ast}{}^D_j+N^A_BQ^{\ast}{}^B_{ij}=Q^{\ast}{}^A_{ij},
\]
which is derived from the identity $N^A_BQ^{\ast}{}^B_{i}=Q^{\ast}{}^A_i$, together with the relation $h^{ij}Q^{\ast}{}^A_iQ^{\ast}{}^B_j=G^{EF}N^A_EN^B_F$. As a result, we get the following representation for $j^{ A}_{\Roman 1}$:
\begin{equation}
 j^{ A}_{\Roman 1}=\frac12h^{BM}N^A_{B, M}+\frac12h^{\tilde B \tilde M}\Bigl(\, {}^{ \mathrm  H}{\tilde \Gamma}^{ A}_{\tilde B\tilde M}-N^A_C\,{}^{ \mathrm  H}{\tilde \Gamma}^{ C}_{\tilde B\tilde M}\Bigr).
\label{j_A_deriv_N}
\end{equation}
 Note that in $j^{A}_{\Roman 1}$ not all $N^A_{\tilde B,\tilde M}$ components are nonzero, because in our case $N^A_b= 0$ and then its derivatives are also equal to zero.

It also follows from $N^a_RQ^{\ast}{}^R_{ij}=N^a_RN^R_{B,D}\,Q^{\ast}{}^B_iQ^{\ast}{}^D_j$  that 
\begin{align}
 j^{ a}_{\Roman 1}&=-\frac12N^{a}_{C}\Bigl(\,h^{BM}N^C_{B, M}+h^{\tilde B\tilde M}\,{}^{ \mathrm  H}{\tilde \Gamma}^{ C}_{\tilde B\tilde M}\Bigr)
 \nonumber\\
 &=\;\;\;\frac12h^{CM}N^a_{C,M}-\frac12N^a_Ch^{\tilde B\tilde M}\,{}^{ \mathrm  H}{\tilde \Gamma}^{ C}_{\tilde B\tilde M}
\label{j_a_deriv_N}
 \end{align}
(the last line follows from differentiating $N^a_CN^C_B=0$ and taking into account $h^{BM}N^C_B=h^{CM}$).
Then we can conclude that the resulting $b^A_{\Roman 1}(Q^{\ast}(x),\tilde f)$ and $b^a_{\Roman 1}(Q^{\ast}( x ) ,\tilde f)$ should be used to determine the drift coefficients of the stochastic differential equation for the stochastic process on $\tilde{\mathcal M}$, which is locally represented by the processes $(Q^ {\ast} { } ^ A(x (t)),\tilde f ^a(t))$.

\subsection{Diffusion coefficients and stochastic differential equations}
The diffusion coefficients of the equation (\ref{sde_Q_ast(x)}) can be obtained if we first  assume that the equality $Q^{\ast}{}^A_i X^i_ {\bar m}dw ^{\bar m} _t = \tilde {\mathscr X}^A_{\bar M}dw^{\bar M} _t$ holds for some matrix $\tilde{\mathscr X}^A_{\bar M}$. Taking the `square' of the equality and using the main property of the Wiener process, according to which  $<dw ^{\bar m}_t,dw ^{\bar n}_t>=\delta^{\bar m\bar n}dt$ and also $<~dw ^{\bar M}_t,dw ^{\bar N}_t>=\delta^{\bar M\bar N}dt$, we find that the matrix $\tilde {\mathscr X}^A_{\bar M}$ satisfies the  local relation $\sum_{\bar M}\tilde{\mathscr X}^A_{\bar M}\tilde{\mathscr X}^B_{\bar M}=G^{CD}N^A_CN^B_D$. This can be verified by using a special representation for the projection operator $N^A_B$: $$N^A_B(Q^{\ast}(x))=G^{\rm H}_{BD}(Q^{\ast}(x)) Q^{\ast}{}^D_ih^{ij}Q^{\ast}{}^A_j.$$

 In turn, the resulting  relation allows us to define matrix ${\mathscr X}^A_{\bar M}$ so that
 $\tilde{\mathscr X}^A_{\bar M}=N^A_C{\mathscr X}^C_{\bar M}$, and hence $\sum_{\bar M}{\mathscr X}^A_{\bar M}{\mathscr X}^B_{\bar M}=G^{AB}$. 

Note that the relations that we have used to derive the diffusion coefficients are possible only because these coefficients in stochastic differential equations with Wiener processes are determined up to orthogonal transformations.

To obtain a stochastic differential equation describing diffusion on a submanifold $\tilde{\mathcal M}$ in terms of the variable given on the external manifold  (i.e., in our case, using dependent coordinates  on charts of the external manifold), it is necessary to redefine the coordinates  $Q^ { \ast }(x(t))$ of local stochastic processes $(Q^{\ast}{}^A(x(t)),\tilde f ^a(t) )$ for new coordinates  $Q^{\ast } ( t ) $\footnote{We denote new stochastic variable by the same letter.}. In addition, we require that the new process, represented by the local processes $(Q^{\ast}{}^A(t),\tilde f ^a(t))$, at the initial moment of time also be given on the submanifold $\tilde {\mathcal M }$.

Thus, taking into account the new representations obtained for the drift and diffusion coefficients, our local stochastic differential equations (\ref{sde_Q1}) and (\ref{sde_f1}) can be rewritten as follows:
\begin{align}
 &dQ^{\ast}{}^A(t)=\mu^2\kappa\Bigl(-\frac12h^{\tilde B \tilde M}\,{}^{ \mathrm  H}{\tilde \Gamma}^{ A}_{\tilde B\tilde M}+j^{ A}_{\Roman 1}+j^{ A}_{\Roman 2}\Bigr)dt
 +\mu\sqrt{\kappa}N^A_C\mathscr X^C_{\bar M}dw^{\bar M}_t,
 \label{sde_Q_ast_j}\\
  &d\tilde f^a(t)=\mu^2\kappa\Bigl(-\frac12h^{\tilde B \tilde M}\,{}^{ \mathrm  H}{\tilde \Gamma}^{ a}_{\tilde B\tilde M}+j^{ a}_{\Roman 1}+j^{ a}_{\Roman 2}\Bigr)dt
  +\mu\sqrt{\kappa}\bigl(N^a_C\mathscr X^C_{\bar M}dw^{\bar M}_t+\mathscr X^a_{\bar b}dw^{\bar b}_t\bigr).
  \label{sde_f_j}
 \end{align}
The terms on the right-hand sides of the above equations now depend on $Q^{\ast}(t)$ and $\tilde f(t)$.
Also note, that in these equations  we have introduced a new notation for the drift coefficients  $b_{\Roman 2}$ defined  in  (\ref{b_2}). Hereinafter they will be denoted as $j_{\Roman 2}$.

 Solutions of the local  equations
    (\ref{sde_Q_ast_j}), (\ref{sde_f_j}), and (\ref{sde_a1}) are the coordinate representatives of the local stochastic process ${\zeta}^{{\tilde{\varphi}}^{\tilde{\cal P}}}(t)=(Q^{\ast}_t{}^A,\tilde f^a_t, a^{\alpha}_t)$ given on a chart of the principal fiber bundle. The set of solutions of such equations on local charts determines the stochastic evolution family of mappings of the the manifold $\tilde{\cal P}$, considered as the total space of the principal fiber bundle $\pi'$. As in \cite{Dalecky_1,Dalecky_2}, these local evolution families of mappings generate a global stochastic evolution family,  
        which  by definition of the cited papers, is a global stochastic process $\zeta_t$ in the principal fiber bundle $\rm P(\tilde{\mathcal M},\mathcal G)$. The local  stochastic process  ${\zeta}^{{\tilde{\varphi}}^{\tilde{\cal P}}}(t)={\tilde{\varphi}}^{\tilde{\cal P}}({\zeta}_t)$ is a local representative of the global stochastic process $\zeta_t$ on the chart of the principal fiber bundle with the coordinate homeomorphism 
        $\tilde{\varphi}^{\tilde{\cal P}}= \tilde \varphi^{-1}\circ \varphi^{\tilde{\cal P}}$.


 We also note that transition from the local stochastic processes $(Q^A_t, f^b_t)$ to the local processes $(Q^{\ast}_t{}^A,\tilde f^a_t, a^{\alpha}_t)$, performed using the mapping $\tilde \varphi^{-1}$, is the phase-space transformation of the stochastic processes.   It is known that such a transformation does not change the probabilities and also the transition probabilities.
 This allows us to rewrite the right-hand side of (8) as the  expectation which is taken over the distribution of the local process ${\zeta}^{{\tilde{\varphi}}^{\tilde{\cal P}}}(t)$:
 \begin{equation*}
 {\rm E}_{s,{\tilde{\varphi}}^{\tilde{\cal P}}(p,v)}[\phi (\,({\varphi}^{\tilde{\cal P}})^{-1}({\tilde{\varphi}}({\zeta}^{{\tilde{\varphi}}^{\tilde{\cal P}}}(t))\,)]={\rm E}_{s,{\tilde{\varphi}}^{\tilde{\cal P}}(p,v)}[\tilde{\phi} ({\zeta}^{{\tilde{\varphi}}^{\tilde{\cal P}}}(t)\,)].
 \end{equation*}
 When deriving such a representation, it was taken into account that 
 $\eta_t^{\varphi^{\tilde{\cal P}}}=\tilde \varphi({\zeta}_t^{{\tilde{\varphi}}^{\tilde{\cal P}}})$ and
 $({\varphi}^{\tilde{\cal P}})^{-1}\circ{\tilde{\varphi}}=({\tilde{\varphi}}^{-1}\circ{\varphi}^{\tilde{\cal P}})^{-1}\equiv ({\tilde{\varphi}}^{\tilde{\cal P}})^{-1}$. Moreover, we have replaced the function $\phi$ under the expectation  sign with the function $\tilde{\phi}=\phi\circ({\tilde{\varphi}}^{\tilde{\cal P}})^{-1}$.

 The global evolution semigroup for the process $\zeta(t)$ is  given by the  limit of the superposition of the local semigroups associated with   local stochastic processes 
${\zeta}^{{\tilde{\varphi}}^{\tilde{\cal P}}}(t)$:
\begin{equation}
\psi _{t_b}(p_a,v_a,t_a)=
{\lim}_q \bigl[{\tilde U}_{{\zeta}^{{\tilde{\varphi}}^{\tilde{\cal P}}}}(t_a,t_1)\cdot\ldots\cdot
{\tilde U}_{{\zeta}^{{\tilde{\varphi}}^{\tilde{\cal P}}}}
(t_{n-1},t_b) 
{\tilde \phi} _0\bigr](Q^{\ast}_a,\tilde f _a, \theta _a),
\label{glob_semigr_zeta}
\end{equation}
where the boundary values of ${\zeta}^{{\tilde{\varphi}}^{\tilde{\cal P}}}(t_a)\equiv(Q^{\ast}_a,\tilde f _a, \theta _a)$ in the right-hand side of the equation (\ref{glob_semigr_zeta}) should be expressed in terms of $(p_a,v_a)$ with the help of inverse transformation $({\tilde{\varphi}}^{\tilde{\cal P}})^{-1}$.

Note that the local semigroups ${\tilde U}_{{\zeta}^{{\tilde{\varphi}}^{\tilde{\cal P}}}}\tilde{\phi}$ are now  defined  as  
\begin{eqnarray}
&&{\tilde U}_{{\zeta}^{{\tilde{\varphi}}^{\tilde{\cal P}}}}(s,t) 
{\tilde \phi} (Q^{\ast}_0,\tilde f_0,\theta _0)={\rm E}_
{s,(Q^{\ast}_0,\tilde f_0,\theta _0)}[
\tilde{\phi}(Q^{\ast}(t),\tilde f(t),a(t))],\;\;
\nonumber\\
&&\,\,s< t,\;\;Q^{\ast}(s)=Q^{\ast}_0,\,\tilde f(s)=\tilde f_0,\,a(s)=\theta _0.
\label{local_semigr_zeta}
\end{eqnarray}

We may consider the global stochastic process $\zeta(t)$ as  consisting  of two components: $\;\;\zeta(t)=\{\xi_{\tilde \Sigma}(t),a_{\mathcal G}(t)\}$. The first process describes a special stochastic evolution (due to a certain form of stochastic differential equations and their initial conditions) on the submanifold ${\tilde\Sigma}$ (on the gauge surface), and the second one describes the evolution on the  orbits of the principal fiber bundle.
Then, taking into account the potential term, the global semigroup  associated with the stochastic process $\zeta(t)$  can be symbolically written in the following form:
  \[
  {\psi}_{t_b} (p_a,v_a,t_a)={\rm E}\Bigl[\tilde{\phi
  }_0({\xi}_{\tilde\Sigma}(t_b),a_{\mathcal G}(t_b))\exp
  \{\frac 1{\mu ^2\kappa m}\int_{t_a}^{t_b}
  \tilde{V}({\xi}_{\tilde\Sigma}(u))du\}\Bigr],
  \]
where ${\xi}_{\tilde\Sigma} (t_a)$ is such a point on ${\tilde\Sigma}$ that has local coordinates $(Q^{\ast}_a,\tilde f_a)$ ,
 and the point $a_{\mathcal G}(t_a)\in {\mathcal G}$  has the local coordinates $\theta _a$. Also,
${{\tilde{\varphi}}^{\tilde{\cal P}}}(p_a,v_a)=(Q^{*}_a,\tilde f_a,\theta _a)$.

\section{Factorization  of the path integral measure}

To perform the  reduction procedure in the path integral, i.e. to transform the original path integral into the path integral discribing the ``quantum evolution'' on the orbit space of the principal bundle, we must first factorize the path integral measure in the original path integral. For dynamical systems with symmetry this can be done by the method developed  in our works 
\cite{Storchak_1,Storchak_98,Storchak_2,Storchak_2019,Storchak_2020}, and  as well as  in \cite{Elworthy}.
This method is based on using the optimal nonlinear filtering stochastic differential equation from the stochastic process theory \cite{Lipcer, Pugachev}.

The equation deals with the evolution of the conditional mathematical expectation of a function that depends on  both an unknown signal process (the process $a_{\mathcal G}(t)$ in our case) and the observation process (the stochastic process $\xi_{\tilde\Sigma}(t)$) relative to the sub-$\sigma $-algebra generated by the observation process. 

 When deriving such an equation in the stochastic theory, it is assumed that the signal process $Z_t$ and the  observation process $Y_t$ satisfy the following stochastic differential equations:
\begin{eqnarray*}
 &&dZ_t=\varphi (Y,Z,t)dt+X(Y,Z,t)\,dw_1(t)+X'(Y,Z,t)\,dw_2(t),\\
&&dY_t=\varphi_1(Y,Z,t) dt+X_1(Y,t) \,dw_2(t),
\end{eqnarray*}
where $w_1(t)$ and $w_2(t)$ are independent Wiener processes.

These equations satisfy two main requirements. The first is that
the diffusion coefficient $X_1$ must not depend on the process $Z_t$. And the second requirement is that  in the equation for the observation process $Y_t$ there should not be a term with $dw_1(t)$.  Note that in our stochastic differential equations, the second requirement will be satisfied  due to the presence of the corresponding projection operators in the diffusion coefficients.

The nonlinear filtering stochastic differential equation for
the conditional mathematical expectation $\hat f=\rm E[f(Y_t,Z_t,t)|{\mathcal Y}^t_{t_0}]$, (${\mathcal Y}^t_{t_0}$ is the sub-$\sigma $-algebra generated by the observation process 
$Y_t$), has the following form:
\begin{eqnarray} 
d\hat f(t)&=&{\rm E}[f_t+f_z\varphi +\frac12 f_{zz}(X X^T)|{\mathcal Y}^t_{t_0}]dt
\nonumber\\
&&\!\!\!+{\rm E}\Bigl[f(\varphi_1-\hat \varphi_1)+f_z(X X^T_1)|{\mathcal Y}^t_{t_0}\Bigr](X_1 X^T_1)^{-1}(dY_t-\hat \varphi_1 dt),
\label{eq_filtr_pugach}
\end{eqnarray}
where $\hat \varphi_1\equiv{\rm E}[\varphi_1(Y_t,Z_t,t)|{\mathcal Y}^t_{t_0}]$.

 By the  properties of the conditional mathematical expectation of Markov processes,
 each local semigroup (\ref{local_semigr_zeta}) of the global semigroup (\ref{glob_semigr_zeta}) can be represented as follows:
  \begin{equation}
{\tilde U}_{{\zeta}^{{\tilde{\varphi}}^{\tilde{\cal P}}}}(s,t) {\tilde
\phi} (Q^{\ast}_0,\tilde f_0,\theta _0)=
{\rm E}
\Bigl[{\rm E}\Bigl[\tilde{\phi }(Q^{\ast}(t),\tilde f(t),a(t))\mid
(
{\cal F}_{(Q^{\ast},\tilde f)})_{s}^{t}\Bigr]\Bigr].
\label{loc_cond_expect}
\end{equation}

 We are interested in the optimal nonlinear filtering equation for the conditional  expectation of $\tilde{\phi }$ with respect to the sub-$\sigma $-algebra  $({\cal F}_{(Q^{\ast},\tilde f)})_{s}^{t}$:
  \[
 \hat{\widetilde{\phi }}(Q^{\ast}(t),\tilde f(t))\equiv 
 {\rm E}\Bigl[\tilde{\phi }(Q^{\ast}(t),\tilde f(t),a(t))\mid (
 {\cal F}_{(Q^{\ast},\tilde f)})_{s}^{t}\Bigr].
\]
Such an equation can be obtained from (\ref{eq_filtr_pugach}) provided that
the process $\xi_{\Tilde\Sigma}$, which is locally determined by the processes  $Q^{\ast}{}^A(t)$ and $\Tilde f^a(t)$, should be considered as an observation process corresponding to the stochastic process $Y_t$  in the above stochastic differential equations. 
Then, in our case,  the local stochastic differential equation for the observation process can be written in the following form: 
\begin{equation}
 {dQ^{\ast}_t{}^A\choose d\tilde f^a_t}=(\mu^2\kappa) {b^A\choose b^a}dt+\mu \sqrt{\kappa}{N^A_C{\mathscr X}_{\bar M}^C\;\; 0\choose {N^a_C{\mathscr X}_{\bar M}^C\;\; {\mathscr X}^a_{\bar b}}} {d w^{\bar M}_t\choose d w^{\bar b}_t}.
\label{sde_Q_f}
\end{equation}
 Note  that the local stochastic  process $(Q^{\ast}{}^A(t),\tilde f^a(t))$  is given  on $\tilde \Sigma$ and is used to describe the stochastic evolution on the base space $\tilde{\mathcal M}$ of the principal fiber bundle 
$\rm {P}(\tilde{\mathcal M},\mathcal G)$. 

Also note that the stochastic process $a^{\alpha}(t)$  satisfying the equation
\[
 da^{\alpha}_t=(\mu^2\kappa)b^{\alpha} dt +\mu \sqrt{\kappa}\Lambda^{\beta}_C{\bar v}^{\alpha}_{\beta}\mathscr X^C_{\bar M}dw^{\bar M}_t
\]
corresponds to the signal  process $Z_t$. Therefore, in our case, instead of the coefficient $X$ of the equation (\ref{eq_filtr_pugach}), we should use $\Lambda^{\beta}_C{\bar v}^{\alpha}_{\beta}{\mathscr X}_{\bar M}^C$, and  $X_1$ must be given by the diffusion matrix of the equation (\ref{sde_Q_f}):  
\begin{equation*}
 \displaystyle
X_1=\left(
\begin{array}{cc}
N^A_C {\mathscr X}_{\bar M}^C & 0\\
N^a_C {\mathscr X}_{\bar M}^C & {\mathscr X}^a_{\bar b}
\end{array}
\right).
\end{equation*}

The remaining coefficients of the equation (\ref{eq_filtr_pugach}), which are necessary for the derivation of the nonlinear filtering stochastic differential equation of our problem, are easy to find, and they
are given by the following expressions:
\[
 X\cdot X^{\rm T}=G^{CD}\Lambda^{\alpha'}_C\Lambda^{\beta'}_D{\bar v}^{\alpha}_{\alpha'}{\bar v}^{\beta}_{\beta'},
\]
\begin{equation*}
\displaystyle 
X\cdot X^{\rm T}_1={\bar v}^{\alpha}_{\beta}\cdot \left (\begin{array}{cc}
 N^B_{C'}\Lambda^{\beta}_CG^{CC'}& N^a_D\Lambda^{\beta}_CG^{CD}\\
0 & 0 \\
\end{array}
\right),
\end{equation*}
\begin{equation*}
 \displaystyle 
X_1\cdot X^{\rm T}_1=\left(
\begin{array}{cc}
G^{CE}N^B_CN^D_E &G^{CD}N^B_CN^a_D\\
G^{EB}N^D_EN^b_B & G^{AB}N^a_AN^b_B +G^{ab}\\
\end{array}
\right),
\end{equation*}

\begin{equation*}
 \displaystyle 
(X_1\cdot X^{\rm T}_1)^{-1}=\left(
\begin{array}{cc}
(P_{\bot})^{A'}_A{\tilde G}^{\rm H}_{A'B'}(P_{\bot})^{B'}_B & (P_{\bot})^{A'}_A{\tilde G}^{\rm H}_{A'a}\\
(P_{\bot})^{B'}_B{\tilde G}^{\rm H}_{bB'} & {\tilde G}^{\rm H}_{ab}\\
\end{array}
\right).
\end{equation*}

 As a result, we get  the following stochastic differential equation of the optimal nonlinear filtering:
\begin{eqnarray}
 &&d \hat{\tilde\phi}(Q^{\ast}(t),\tilde f(t))=\mu^2\kappa\Bigr\{-\frac12\Bigl[d^{-1/2}H^{-1/2}\frac{\partial}{\partial Q^{\ast}{}^A}(d^{1/2}H^{1/2}G^{EF}N^A_EN^C_F\,\underset{\scriptscriptstyle{(\gamma)}}{{\mathscr A}^{\beta}_C})
\nonumber\\
&&+G^{EC}{\Lambda}^{\beta}_E{\Lambda}^{\mu}_C\,d^{-1/2}H^{-1/2}\frac{\partial}{\partial \tilde f^b}(d^{1/2}H^{1/2}K^b_{\mu})
\Bigr]{\rm E}\bigl[\bar L_{\beta}{\tilde\phi}(Q^{\ast}_t,\tilde f_t,a_t)|(\mathcal F_{Q^{\ast},\tilde f})^t_s\bigr]
\nonumber\\
&&+\frac12(G^{BC}{\Lambda}^{\alpha'}_B{\Lambda}^{\beta'}_C){\rm E}\bigl[\bar L_{\alpha'}\bar L_{\beta'}{\tilde\phi}(Q^{\ast}_t,\tilde f_t,a_t)|(\mathcal F_{Q^{\ast},\tilde f})^t_s\bigr]\Bigr\}dt
\nonumber\\
&&+\mu\sqrt{\kappa}\,{\Lambda}^{\beta}_C{\tilde \Pi}^C_E {\mathscr X}^E_{\bar M}{\rm E}\bigl[\bar L_{\beta}{\tilde\phi}(Q^{\ast}_t,\tilde f_t,a_t)|(\mathcal F_{Q^{\ast},\tilde f})^t_s\bigr]dw^{\bar M}_t
\nonumber\\
&&+\mu\sqrt{\kappa}\,{\Lambda}^{\beta}_C{\tilde \Pi}^C_a {\mathscr X}^a_{\bar b}{\rm E}\bigl[\bar L_{\beta}{\tilde\phi}(Q^{\ast}_t,\tilde f_t,a_t)|(\mathcal F_{Q^{\ast},\tilde f})^t_s\bigr]dw^{\bar b}_t.
\label{eq_filtr}
\end{eqnarray}
To simplify this equation, we  apply the Peter-Weyl theorem to the function $\tilde\phi$, considered as a function given on a group $\mathcal G$.

According to this theorem, the function $\tilde \phi$ can be represented as
$$\tilde \phi(Q^{\ast},\tilde f,a)=\sum_{\lambda,p,q}c^{\lambda}_{pq}(Q^{\ast},\tilde f)D^{\lambda}_{pq}(a),$$
 where $D^{\lambda}_{pq}(a)$\footnote{We have now introduced a different notation for the matrix elements of an irreducible representation to distinguish them from those used previously.} are 
    the matrix elements  of an irreducible representation
    $T^{\lambda}$ of a group $\cal G$:
 $\sum_qD_{pq}^\lambda(a)D_{qn}^\lambda (b)=D_{pn}^\lambda (ab)$.
 
It follows from the properties of conditional mathematical expectations that
\begin{eqnarray*}
{\rm E}\bigl[\tilde \phi(Q^{\ast}(t),\tilde f(t),a(t))|(\mathcal F_{Q^{\ast},\tilde f})^t_s\bigr]&=&\sum_{\lambda,p,q}c^{\lambda}_{pq}(Q^{\ast}(t),\tilde f(t))\,{\rm E}\bigl[D^{\lambda}_{pq}(a(t))|(\mathcal F_{Q^{\ast},\tilde f})^t_s\bigr]\\
&\equiv& \sum_{\lambda,p,q}c^{\lambda}_{pq}(Q^{\ast}(t),\tilde f(t))\,\hat D^{\lambda}_{pq}(Q^{\ast}(t),\tilde f(t)),
\end{eqnarray*}
where
\[
c_{pq}^\lambda (Q^{\ast}(t),\tilde f(t))=d^\lambda \int_{\mathcal
G}\tilde{\phi }
(Q^{*}(t),\tilde f(t),\theta ) 
{\bar D}_{pq}^\lambda (\theta )d\mu (\theta ),
\]
$d^{\lambda}$ is a dimension of an irreducible representation
and 
$d\mu (\theta )$ is a normalized\\ ($\int_{\mathcal G}d\mu
(\theta )=1$)
invariant Haar measure on a group $\mathcal G$.

Then, the equation (\ref{eq_filtr}) is reduced to the stochastic differential equation for
 the  conditional mathematical expectation $\hat{D}_{pq}^\lambda(Q^{\ast}(t),\tilde f(t))$:
\begin{eqnarray}
&&d\hat{D}_{pq}^\lambda =\mu^2\kappa\bigl(\Gamma _1^\beta \,(J_\beta )_{pq^{\prime
}}^\lambda \hat{D}_{q^{\prime }q}^\lambda dt
+\Gamma _2^{\alpha \beta }\,(J_\alpha )_{pq^{\prime }}^\lambda (J_\beta
)_{q^{\prime }q^{\prime \prime }}^\lambda \hat{D}_{q^{\prime \prime
}q}^\lambda \bigr)dt
\nonumber\\
&&+\mu\sqrt{\kappa}(J_\beta )_{pq^{\prime }}^\lambda \hat{D}_{q^{\prime }q}^\lambda\,\Lambda^{\beta}_C \bigl(
{\tilde \Pi}^C_E {\mathscr X}^E_{\bar M} dw^{\bar M}_t+{\tilde \Pi}^C_a {\mathscr X}^a_{\bar b} dw^{\bar b}_t\bigr).
\label{eq_filtr_D}
\end{eqnarray}
In this equation   $(J_\beta )_{pn}^\lambda$ are the infinitesimal generators
of the representation $D^{\lambda}(a)$. They are defined as  
$(J_\beta )_{pq}^\lambda \equiv ({\partial D_{pq}^\lambda
(a)}/{\partial
a^\beta })|_{a=e}$, and we also have
\[
\bar{L}_{\beta} D_{pq}^{\lambda} (a)=\sum\nolimits_{q^{\prime}}(J_{\beta})_{pq^{\prime}}^{\lambda} D_{q^{\prime }q}^\lambda (a).
\]
 We do not write out explicit expressions for the coefficients
 $ \Gamma
_1^{\beta} $ and $ \Gamma_2^{\alpha \beta} $, since they can be easily obtained using the equation (\ref{eq_filtr}).

Also note that  the conditional expectations
  $\hat{D}_{pq}^{\lambda} (Q^{\ast}(t),\tilde f(t))$ also depend 
 on the initial points
  $Q^{\ast}_0=Q^{\ast}(s)$, $\tilde f^a_0=\tilde f^a(s)$ and $\theta^{\alpha}_0=a^{\alpha}(s)$. To avoid awkward notation, we have omitted this dependence in $\hat{D}_{pq}^{\lambda} $.

 The solution of the linear matrix stochastic differential equation (\ref{eq_filtr_D}) can be written \cite{Dalmulti,Stroock} as 
 \begin{equation}
\hat{D}_{pq}^\lambda (Q^{\ast}(t),\tilde f(t))=(\overleftarrow{\exp })_{pn}^\lambda
(Q^{\ast}(t),\tilde f(t),t,s)\,{\rm E}\bigl[D_{nq}^\lambda (a(s))\mid ({\cal F}
_{Q^{\ast},\tilde f})_{s}^t\bigr],
\label{solut_eq_filtr_D}=
\end{equation}
where 
\begin{eqnarray}
&&(\overleftarrow{\exp })_{pn}^\lambda (Q^{\ast}(t),\tilde f(t),t,s)=
\overleftarrow{\exp }%
\int_{s}^t\Bigl\{{\mu}^2\kappa\Bigl[\frac 12d^{\alpha \nu
}(Q^{\ast}(u),\tilde f(u))(J_\alpha
)_{pr}^\lambda (J_\nu )_{rn}^\lambda \Bigr.\Bigr.
\nonumber\\
&&-\Bigl.\Bigl.\frac 12\frac 1{\sqrt{d\, H}}\frac \partial {\partial Q^{\ast}{}^A}\left( \sqrt{d\, H}%
G^{EF}N^A_EN^C_F\underset{\scriptscriptstyle{(\gamma)}}{{\mathscr A}_C^\nu}(Q^{\ast}(u)) \right) (J_\nu )_{pn}^\lambda  
\nonumber\\
&&-\frac12(G^{EC}\Lambda^{\nu}_E\Lambda^{\mu}_C)\frac 1{\sqrt{d\, H}}\frac \partial {\partial \tilde f^b}\left(\sqrt{d\, H}K^b_{\mu}\right)(J_\nu )_{pn}^\lambda \Bigr]du
\nonumber\\
&&+\mu\sqrt{\kappa}\Lambda^{\beta}_C(Q^{\ast}(u)) (J_\beta )_{pn}^\lambda \Bigl[ 
{\tilde \Pi}^C_E {\mathscr X}^E_{\bar M}(Q^{\ast}(u)) dw^{\bar M}(u)+{\tilde \Pi}^C_a {\mathscr X}^a_{\bar b} dw^{\bar b}(u)\Bigr]\Bigr\}
\nonumber\\
&&(H,d,{\tilde \Pi}^C_E, {\tilde \Pi}^C_a\;{\rm depend}\,{\rm on}\;Q^{\ast}(u)\;{\rm and}\,\tilde f(u)) 
\label{multipl_exp}
\end{eqnarray}
is the multiplicative stochastic integral. This integral is defined as a limit of
the sequence of time--ordered multipliers that have been
obtained as a
result of the piecewise breaking of the time interval $[s,t]$, $[s=t_0\le t_1
\ldots \le t_n=t]$.
In (\ref{multipl_exp}), the time order of these
multipliers is indicated by the arrow directed 
to the multipliers given at greater times.

Using the solution  of the matrix stochastic differential equation (\ref{eq_filtr_D}) defined by (\ref{solut_eq_filtr_D}) and (\ref{multipl_exp}), the local semigroup (\ref{loc_cond_expect}) can be represented as follows:
\begin{eqnarray}
&&{\tilde U}_{{\zeta}^{{\tilde{\varphi}}^{\tilde{\cal P}}}}(s,t) {\tilde \phi} (Q^{\ast}_0,\tilde f_0,\theta _0)\nonumber\\
&&\;\;\;\;=\sum_{\lambda ,p,q,q^{\prime }}\!\!{\rm E}
\bigl[
 c_{pq}^\lambda (Q^{\ast}(t),\tilde f(t))
(\overleftarrow{\exp })_{pq^{\prime }}^\lambda 
(Q^{\ast}(t),\tilde f(t),t,s)\bigr] D_{q^{\prime}q}^\lambda (\theta _0).
\label{local_semigr_exp}
\end{eqnarray}
Note that in this representation, the last multiplier on the right is due to
\[
{\rm E}\bigl[D_{nq}^\lambda (a(s))\mid ({\cal F}_{Q^{\ast},\tilde f})_{s}^t\bigr]
=D_{nq}^\lambda (a(s))=D_{nq}^\lambda (\theta _0).
\]
The global evolution semigroup (\ref{glob_semigr_zeta}) is now obtained in accordance with \cite{Dalecky_1,Dalecky_2} as the limit (under the refinement of the subdivision of the time interval $[t_a,t_b]$) of the superposition of the local semigroups that are similar to (\ref{solut_eq_filtr_D}).
We will write the global semigroup in the following symbolic form:
\begin{eqnarray}
{\psi}_{t_b}(p_a,v_a,t_a)
&=&\sum_{\lambda ,p,q,q^{\prime }}{\rm E}
\bigl[
 c_{pq}^\lambda (\xi_{\tilde \Sigma}(t_b))
(\overleftarrow{\exp })_{pq^{\prime }}^\lambda 
(\xi_{\tilde \Sigma}(t),t_b,t_a)\bigr] D_{q^{\prime}q}^\lambda (\theta _{a}),
\nonumber\\
&&\;\;\;\;\;\;\;\;\;(\xi_{\tilde \Sigma}(t_a)=\pi'_{(\tilde{\Sigma})}(p_a,v_a)),
\label{glob_semigr_ksi}
\end{eqnarray}
where the global process $\xi_{\tilde \Sigma} (t)=(\xi_1(t),\xi_2(t))$ is defined  on the submanifold $\tilde \Sigma$.
The    process
$\xi_{\tilde \Sigma} (t)$ is described locally by the stochastic equations (\ref{sde_Q_f}).

Thus,  we have obtained that our initial path integral (\ref{orig_path_int}) is represented  as the sum of the matrix semigroups (the path integrals) given  on the submanifold $\tilde{ \Sigma}$. 
 The coordinate representation of the differential generator (the Hamilton operator) of these matrix semigroups is
 \begin{align}
  &\frac12(\mu^2\kappa)\Bigl\{\Bigr[h^{AB}\frac{\partial^2}{\partial Q^{\ast}{}^A\partial Q^{\ast}{}^B}+2h^{Aa}\frac{\partial^2}{\partial Q^{\ast}{}^A\partial \tilde f^a}
+h^{ab}\frac{\partial^2}{\partial \tilde f^a\partial \tilde f^b}
\nonumber\\
&-\bigl(h^{\tilde B \tilde M}\,{}^{ \mathrm  H}{\tilde \Gamma}^{ A}_{\tilde B\tilde M}-2(j^{ A}_{\Roman 1}+j^{ A}_{\Roman 2})\bigr)\frac{\partial}{\partial Q^{\ast}{}^A}
-\bigl(h^{\tilde B \tilde M}\,{}^{ \mathrm  H}{\tilde \Gamma}^{ a}_{\tilde B\tilde M}
-2(j^{ a}_{\Roman 1}+j^{ a}_{\Roman 2})\bigr)\frac{\partial}{\partial \tilde f^a}\Bigr](I)^{\lambda}_{pn}
\nonumber\\
&
+2N^A_BG^{EB}{\Lambda}^{\beta}_E(J_{\beta})^{\lambda}_{pn}\frac{\partial}{\partial Q^{\ast}{}^A}-2d^{\beta\nu}K^b_{\nu}(J_{\beta})^{\lambda}_{pn}\frac{\partial}{\partial \tilde f^b}
\nonumber\\
&-\Bigl.\Bigl.\Bigl(\frac 1{\sqrt{d\, H}}\frac \partial {\partial Q^{\ast}{}^A}\left(\! \sqrt{d\, H}%
h^{AC}\!\underset{\scriptscriptstyle{(\gamma)}}{{\mathscr A}_C^\nu} \right)  
+G^{EC}{\Lambda}^{\nu}_E{\Lambda}^{\mu}_C\frac{ 1}{\sqrt{d\, H}}\frac{\partial }{\partial \tilde f^b}\left(\sqrt{d\, H}K^b_{\mu}\right)\Bigr)(J_\nu )_{pn}^{\lambda} 
\nonumber\\
&+G^{CD}{\Lambda}^{\alpha}_C{\Lambda}^{\beta}_D(J_{\alpha} )_{pq}^{\lambda}
(J_{\beta} )_{qn}^{\lambda}
\Bigr\},
\label{dif_gen_ksi}
\end{align}
where $(I^\lambda )_{pq}$ is a unity matrix. 

The operator (\ref{dif_gen_ksi}) acts in the space of the sections 
$\Gamma ({\tilde{ \Sigma}},V^{*})$ of the associated covector 
bundle (we consider the backward Kolmogorov equation). 
The scalar product in the space of the sections of 
the associated co-vector  bundle is given by
\begin{equation}
(\psi _n,\psi _m)=\int_{\tilde{ \Sigma}}\langle \psi _n,\psi _m{\rangle}_
{V^{\ast}_{\lambda}}
 \,d^{1/2}\,dv_{\tilde{\Sigma}},
\label{33}
\end{equation}
 where $d=\det (d_{\alpha\beta})$ and  $dv_{\tilde{\Sigma}}$ is the Riemannian volume element on the submanifold ${\tilde{\Sigma}}$. In  local coordinates, it can be presented as $$dv_{\tilde{\Sigma}}(Q^{\ast},\tilde f)=H^{1/2}(Q^{\ast},\tilde f)\,dQ^{\ast}{}^1...dQ^{\ast}{}^{n_
 {\cal P}}d\tilde f^1...\tilde f^{n_{\cal \mathcal V}}.$$ 

To express the matrix semigroup under the  sign of sum in (\ref{glob_semigr_ksi}) in terms of the original semigroup defined on $\tilde{\mathcal P}$, it is necessary to inverse this equality. 
In our previous papers \cite{Storchak_98,Storchak_2}, such an inversion of the analogous equality  was made for the kernels of the corresponding  local semigroups.
To do this, it was assumed that all the necessary restrictions  for the existence of semigroup kernels were fulfilled. 
This method is applicable to the case we are now considering, if we take into account
     local isomorphism of the principal bundle ${\rm P}(\tilde{ \mathcal M},\mathcal G)$ with the trivial bundle $\tilde\Sigma\times \mathcal G\to \tilde\Sigma$. This isomorphism, as shown in \cite{Mitter-Viallet}, leads to the existence of the  relationship  between the charts of these principal fiber bundles. 
As a result, it becomes possible to reverse the relation between local kernels (local Green's functions) which are defined on charts of the principal bundle ${\rm P}(\tilde \Sigma,\mathcal G)$.

If the global semigroup on the left-hand side of (\ref{glob_semigr_ksi}) can be presented as 
\begin{equation}
 {\psi}_{t_b}(p_a,v_a,t_a)=\int G_{\tilde{\mathcal P}}(p_b,v_b,t_b;p_a,v_a,t_a)\phi_0(p_b,v_b)dv_{\tilde{\mathcal P}}(p_b,v_b),
\label{global_semigr_P}
\end{equation}
then comparing the local expression of the right-hand side of (\ref{glob_semigr_ksi}) (i.e. what is obtained  on the charts of the principal fiber bundle  ${\rm P}(\tilde \Sigma,\mathcal G)$ ) with the local expression $G_{\tilde{\mathcal P}}$ given on the corresponding charts from the atlas of the manifold ${\tilde{\mathcal P}}$ (with account of a local isomorphism of the bundles), one can find the following relation between the local Green's functions: 
\begin{eqnarray*}
&&\int_{\mathcal G} G_{\tilde{\mathcal P}}(\alpha_b, F(Q^{\ast}_b,{\theta}_b),\bar D({\theta}_b)\tilde f_b,t_b;\beta _a, F(Q^{\ast}_a,{\theta}_a),\bar D({\theta}_a)\tilde f_a,t_a)
\nonumber\\
&&\;\;\;\;\;\;\;\;\;\times D^{\lambda}_{pq}({\theta}_b)d{\mu}({\theta}_b)=\sum_{q'}G^{\lambda}_{q'p}({\alpha}_b,Q^{\ast}_b,\tilde f_b,t_b;{\beta}_a, Q^{\ast}_a,\tilde f_a,t_a)D^{\lambda}_{q'q}({\theta}_a),
\end{eqnarray*}
where $\alpha_b$ and $\beta _a$ are the labels of the charts.
 Note that to obtain the above relation, $\phi_0(p_b, v_b)$ in (\ref{global_semigr_P}) was also expanded into a series using the Peter-Weyl theorem. This was done after introducing on $\tilde {\mathcal P}$  the bundle coordinates. 

The obtained  relation between the local Green's functions can be reversed:
\begin{eqnarray}
 &&G^{\lambda}_{mn}({\alpha}_b,Q^{\ast}_b,\tilde f_b,t_b;{\beta}_a, Q^{\ast}_a,\tilde f_a,t_a)
\nonumber\\
&&\;\;\;\;\;\;\;\;\;\;=\int_{\mathcal G}G_{\tilde{\mathcal P}}(\alpha_b, Q^{\ast}_b,\tilde f_b,{\theta},t_b;\beta _a, Q^{\ast}_a,\tilde f_a, e,t_a)D^{\lambda}_{nm}({\theta})d{\mu}({\theta}), 
\label{interrelat_green_fun}
\end{eqnarray}
where  $e$ is the unity element of the group $\mathcal G$. Also note that there is the following invariance property:
\begin{eqnarray*}
 &&G_{\tilde{\mathcal P}}(\alpha_b, Q^{\ast}_b,\tilde f_b,{\theta}_b,t_b;\beta _a, Q^{\ast}_a,\tilde f_a, {\theta}_a,t_a)
\nonumber\\
&&\;\;\;\;\;\;\;\;\;\;\equiv G_{\tilde{\mathcal P}}(\alpha_b, F(Q^{\ast}_b,{\theta}_b),\bar D({\theta}_b)\tilde f_b,t_b;\beta _a, F(Q^{\ast}_a,{\theta}_a),\bar D({\theta}_a)\tilde f_a,t_a).
\end{eqnarray*}
To extend the resulting equality (\ref{interrelat_green_fun}) from local charts to the whole manifold, we need to glue
   these local Green's functions. In case of the trivial principal fiber bundle this can be done  using the transition coordinate functions from  the manifold atlas. As a result, we obtain the global integral relation between the  Green's functions:
\begin{eqnarray}
&&G^{\lambda}_{mn}(\pi'_{(\tilde{\Sigma})}(p_b,v_b),t_b;
 \pi'_{(\tilde{\Sigma})}(p_a,v_a),t_a)
 \nonumber\\
&&\;\;\;\;\;\;\;\;\;\;\;\;=\int _{\cal G}G_{\tilde{\cal P}}(p_b\theta,v_b\theta,t_b;
p_a,v_a,t_a) 
D_{nm}^\lambda (\theta )d\mu (\theta ),
\label{green_funk_relat}
\end{eqnarray}
The path integral for the Green's function $G^{\lambda}_{mn}$ can be written symbolically as
\begin{eqnarray}
&&G^{\lambda}_{mn}(\pi'_{(\tilde{\Sigma})}(p_b,v_b),t_b;
 \pi'_{(\tilde{\Sigma})}(p_a,v_a),t_a)=
\nonumber\\
&&{\tilde {\rm E}}_{{\xi_{\tilde \Sigma} (t_a)=\pi'_{(\tilde{\Sigma})} (p_a,v_a)}\atop  
{\xi_{\tilde \Sigma} (t_b)=\pi'_{(\tilde{\Sigma})} (p_b,v_b)}}
\Bigl[(\overleftarrow{\exp })_{mn}^\lambda 
(\xi_{\tilde \Sigma}(t),t_b,t_a)
\exp \{\frac 1{\mu ^2\kappa m}\int_{t_a}^{t_b}
\tilde{V}(\xi_1(u),\xi_2(u))du\}\Bigr]
\nonumber\\
&&=\int\limits_{{\xi_{\tilde \Sigma} (t_a)=\pi'_{(\tilde\Sigma)} (p_a,v_a)}\atop  
{\xi_{\tilde \Sigma} (t_b)=\pi'_{(\tilde{\Sigma})} (p_b,v_b)}} d{\mu}^{\xi_{\tilde \Sigma}}
\exp \{\frac 1{\mu ^2\kappa
m}\int_{t_a}^{t_b}\tilde{V}(\xi_{\tilde \Sigma}(u))du\}
\nonumber\\
&&\times
\overleftarrow{\exp }%
\int_{t_a}^{t_b}\Bigl\{{\mu}^2\kappa\Bigl[\frac 12d^{\alpha \nu
}(\xi_{\tilde \Sigma}(u))(J_\alpha
)_{mr}^\lambda (J_\nu )_{rn}^\lambda \Bigr.\Bigr.
\nonumber\\
&&-\Bigl.\Bigl.\frac 12\frac 1{\sqrt{d\, H}}\frac \partial {\partial Q^{\ast}{}^A}\left( \sqrt{d\, H}%
G^{EF}N^A_EN^C_F\underset{\scriptscriptstyle{(\gamma)}}{{\mathscr A}_C^\nu} \right) (J_\nu )_{mn}^\lambda  
\nonumber\\
&&-\frac12(G^{EC}\Lambda^{\nu}_E\Lambda^{\mu}_C)\frac 1{\sqrt{d\, H}}\frac \partial {\partial \tilde f^b}\left(\sqrt{d\, H}K^b_{\mu}\right)(J_\nu )_{mn}^\lambda \Bigr]du
\nonumber\\
&&-\mu\sqrt{\kappa}{\Lambda}^{\beta}_C(J_\beta )_{mn}^\lambda\Bigl[ 
{\tilde \Pi}^C_E {\mathscr X}^E_{\bar M}({\xi}_1(u)) dw^{\bar M}(u)+{\tilde \Pi}^C_a {\mathscr X}^a_{\bar b} dw^{\bar b}(u)\Bigr]\Bigr\}.
\label{path_int_G_mn}
\end{eqnarray}
The semigroup with this kernel acts in the space of the equivariant functions given on $\tilde{\mathcal P}$:
\[
 \tilde\psi_n(pg,vg)= D^{\lambda}_{mn}(g)\tilde\psi_m(p,v).
\]
In local coordinates, the isomorphism of these functions with the functions $\psi_n\in  \Gamma ({{\tilde \Sigma}},V^{*})$
is represented as follows:
\[
 \tilde\psi_n(F(Q^{\ast},e),\bar D^b_c(e)\tilde f^c)=\psi_n(Q^{\ast},\tilde f).
\]
The semigroup with the Green's function $G^{\lambda}_{mn}$ is used to describe the ``quantum evolution'' of the reduced dynamical system on the manifold $\tilde{\mathcal M}$ -- the orbit space of the principal fiber bundle.
This path integral reduction  corresponds to the case of the reduction of a mechanical system with symmetry
to a nonzero momentum level \cite{AbrMarsd}.

Note  that in our case, in order to describe the evolution  on the orbit space $ \tilde{\mathcal M} $ we use an additional surface  $\tilde {\Sigma}$ (`gauge surface') on which diffusion is given locally by the solution of the local stochastic differential equation (\ref{sde_Q_f}). In fact, this equation is represented by two equations:  (\ref{sde_Q_ast_j}) and (\ref{sde_f_j}). 
We see that in both of these equations, in the drift coefficients, there are terms that we denoted as $j_{\Roman 1}$ and $j_{\Roman 2}$.    The presence of $j_{\Roman 1}$-terms in the drift coefficients of our  stochastic equation on $\tilde {\Sigma}$ is essential for the correct description of the stochastic process  on the manifold $ \tilde{{\Sigma}} $, considered as a submanifold in the ambient manifold with the horizontal metric.
On the other hand, the $j_{\Roman 2}$-terms in the drift coefficients of stochastic differential equations are not necessary.  
  
Transformation of a measure in a path integral by replacing the stochastic process ${\xi}_{\tilde{\Sigma}}$, determined by solutions of local stochastic differential equations (\ref{sde_Q_f}), with the process ${ \tilde\xi}_{ \tilde{\Sigma}}$, described by the same local stochastic equations, but without the $j_{\Roman 2}$ terms,\footnote{For the components of the local stochastic process representing the global stochastic process ${\tilde{\xi}_{\tilde\Sigma}}(t)$ on charts, we will use the same notation as before, namely $(Q^{\ast}_t{}^A,\tilde f_t^a)$.}
can be performed using the Girsanov-Cameron-Martin transformation.
 
In our article, we consider such a transformation  for the  reduction onto a  zero-momentum level. 
In this case, i.e. when $\lambda =0$, instead of (\ref{green_funk_relat}), we  get the integral relation between the scalar Green's functions (semigroup kernels) acting in the spaces of scalar functions given on $\tilde{\mathcal P}$ and on $\tilde{\Sigma}$. 

Note, that  the standard Girsanov-Cameron-Martin formula for the path integral measure transformation deals with the case when the matrix representing diffusion coefficient of the stochastic differential equation is nondegenerate. The elements of such a matrix in (\ref{sde_Q_f}) are defined using   projection operators. Therefore, this matrix is  degenerate. In spite of this, an analogue of the Girsanov-Cameron-Martin formula can also be derived in our case. Its derivation  (using the It\^{o}'s differentiation formula)  follows from the assumption that the solution of a parabolic differential equation with the differential operator (\ref{dif_gen_ksi}) (taken at $\lambda =0$) is unique (modulo those ambiguities that we have in the problem under consideration).

 The Radon-Nikodym derivative of the measure 
${\mu}^{{\xi}_{\tilde\Sigma}}$  with respect to the measure
${\mu}^{\tilde{\xi}_{\tilde\Sigma}}$ is defined as
\begin{equation}
\frac{d{\mu}^{{\xi}_{\tilde\Sigma}}}
{d{\mu}^{\tilde{\xi}_{\tilde\Sigma}}}
({\tilde{\xi}_{\tilde\Sigma}}(t))
=\exp\int^t_{t_a}\Bigl[\mu^2\kappa<A^{-1}\!\cdot \vec{j}{}_{\Roman 2},\,d\bar{w}_s>-\frac12(\mu^2\kappa)^2||A^{-1}\!\cdot\vec{j}{}_{\Roman 2}||^2 ds\Bigr], 
\label{girs}
\end{equation}
where $A^{-1}$ is the matrix which is (pseudo)inverse to the `diffusion' matrix $X_1$ of the local stochastic differential equation (\ref{sde_Q_f}), the Wiener process $\bar{w}_s$ consists of two independent Wiener processes:  $\bar{w}_s=(w_s^{\bar M}, w_s^{\bar b})$, and $\vec{j}{}_{\Roman 2}$ is  the projection  of the mean curvature vector field  of the orbit onto the submanifold $\tilde{\Sigma}$. It has  the following components:
\begin{eqnarray}
&& j_{\Roman 2}^A=-\frac12 \,G^{CC'}N^A_{C'}N^{B'}_CG_{BB'}\,d^{\alpha \beta}\bigl(\nabla_{K_{\alpha}}K_{\beta}\bigr)^B,
\nonumber\\
&&j_{\Roman 2}^a=-\frac12d^{\alpha \beta}\Bigl(N^a_B\,\bigl(\nabla_{K_{\alpha}}K_{\beta}\bigr)^B+\bigl(\nabla_{K_{\alpha}}K_{\beta}\bigr)^a\Bigr).
\label{j_2}
\end{eqnarray}

To get an explicit representation of $A^{-1}\!\cdot \vec{j}{}_{\Roman 2}$ in (\ref{girs}), we first rewrite the terms on the right-hand side of the equations (\ref{j_2}) using the following identities:
\begin{eqnarray*}
 &&d^{\alpha \beta}\bigl(\nabla_{K_{\alpha}}K_{\beta}\bigr)^B=-\frac12\Bigl(G^{BC}N^A_C{\sigma}_A+G^{BC}N^a_C\,{\sigma}_a\Bigr),
\nonumber\\
&&d^{\alpha \beta}\bigl(\nabla_{K_{\alpha}}K_{\beta}\bigr)^a=-\frac12 G^{aq}{\sigma}_q,
\end{eqnarray*}
where ${\sigma}_A=d^{\alpha \beta}\frac{\partial }{\partial Q^{\ast A}}d_{\alpha \beta}\equiv \frac{\partial }{\partial Q^{\ast A}}\ln d$ and ${\sigma}_a=\frac{\partial }{\partial {\tilde f}^a} \ln d,$   $d=\det d_{\alpha \beta}$.

As a result, we obtain 
\begin{eqnarray}
\displaystyle
\Bigl(
\begin{array}{cc} 
j_{\Roman 2}^A\\
j_{\Roman 2}^a
\end{array}
\Bigr)
&=&\frac14 
\Bigl(
\begin{array}{cc}
G^{BD}N^A_{B}N^{C}_D & G^{BD}N^A_{B}N^{b}_D\\
G^{BD}N^a_{B}N^{C}_D & G^{BD}N^a_{B}N^{b}_D+G^{ab}\\
\end{array}\Bigr)
\Bigl(
\begin{array}{cc}
 {\sigma}_C\\ 
{\sigma}_b\\
\end{array}
\Bigr)
\nonumber\\
&=&\frac14\Bigl(X_1\cdot X_1^{\top}\Bigr)\Bigl(
\begin{array}{cc}
 {\sigma}_C\\ 
{\sigma}_b\\
\end{array}
\Bigr).
\label{j_2_sigma}
\end{eqnarray}
In turn, this means that
\[
\displaystyle A^{-1}\!\cdot \vec{j}{}_{\Roman 2}=X_1^{\top}\cdot \vec{j}{}_{\Roman 2}=
\frac14 
\Bigl(
\begin{array}{cc}
N^A_C\mathscr{X}^C_{\bar M}  & N^b_D\mathscr{X}^D_{\bar M}\\
0 & \mathscr{X}^b_{\bar b}\\
\end{array}\Bigr)
\Bigl(
\begin{array}{cc}
 {\sigma}_C\\ 
{\sigma}_b\\
\end{array}
\Bigr).
\]
Thus, the terms under the integral in (\ref{girs}) have the following form:
\[
 <A^{-1}\!\cdot \vec{j}{}_{\Roman 2},\,d\bar{w}_s>=\frac14 \bigl [(N^A_C\mathscr{X}^C_{\bar M}{\sigma}_A+N^a_D\mathscr{X}^D_{\bar M}{\sigma}_a)dw^{\bar M}_s+\mathscr{X}^a_{\bar b}dw^{\bar b}_s\bigr]
\]
and
\begin{align*}
&||A^{-1}\!\cdot\vec{j}{}_{\Roman 2}||^2
\nonumber\\
&\;=\frac{1}{16}\bigl[G^{CD}N^A_CN^B_D{\sigma}_A {\sigma}_B+2G^{CD}N^a_CN^B_D{\sigma}_a {\sigma}_B+(G^{CD}N^a_CN^b_D+G^{ab}){\sigma}_a {\sigma}_b\bigr]\\
&\;=\frac{1}{16}\bigl[h^{AB}{\sigma}_A {\sigma}_B+2h^{aB}{\sigma}_a {\sigma}_B+h^{ab}{\sigma}_a {\sigma}_b\bigr].
\end{align*}
For brevity, the quadratic form in square brackets of the above expression will be further denoted as $<\partial\sigma,\partial\sigma>_{\tilde {\Sigma}}$.

The argument of the exponential  on the right-hand  side of the equation (\ref{girs}) is represented by the sum of the stochastic and ordinary integrals. This exponential function can be rewritten to include only ordinary integrals.
This can be done using the special identity (known as the It\^{o}'s identity). 

In our case, the identity is obtained from the solution of a local stochastic differential equation having $\exp\sigma(Q^{\ast}_t{}^A,\tilde f_t^a)$ as an unknown function. The equation itself is derived by taking the stochastic It\^o differential of this function, provided that now the components $Q^{\ast}_t{}^A$ and $\tilde f_t^a$ of the local stochastic process representing  the global stochastic process ${\tilde{\xi}_{\tilde\Sigma}}(t)$ must satisfy the stochastic differential equations (\ref{sde_Q_ast_j}) and (\ref{sde_f_j}) without ${j}{}_{\Roman 2}$ terms.

Performing calculations, one can find the following identity:
\begin{align*}
&\exp\int^{{t}}_{t_{a}}(\mu\sqrt{\kappa})\Bigl[ (N^A_C\mathscr{X}^C_{\bar M}{\sigma}_A+N^a_D\mathscr{X}^D_{\bar M}{\sigma}_a)dw^{\bar M}_s+\mathscr{X}^a_{\bar b}dw^{\bar b}_s\Bigr]
\nonumber\\
&
=\left(\frac{exp(\sigma(Q^{\ast}(t),\tilde f(t))}{exp(\sigma(Q^{\ast}(t_a),\tilde f(t_a))}\right)
\exp \biggl\{\int^{{t}}_{t_{a}}\frac12(\mu^2{\kappa})\Bigl[h^{AB}\sigma_{AB}+2h^{Ab}\sigma_{Ab}
+h^{ab}\sigma_{ab}
\nonumber\\
&
-\Bigl(h^{\tilde B \tilde M}\,{}^{ \mathrm  H}{\tilde \Gamma}^{ A}_{\tilde B\tilde M}-2j^{ A}_{\Roman 1}\Bigr)\sigma_A
-\Bigl(h^{\tilde B \tilde M}\,{}^{ \mathrm  H}{\tilde \Gamma}^{ a}_{\tilde B\tilde M}-2j^{ a}_{\Roman 1}\Bigr)\sigma_a\Bigr]ds\biggr\}.
\end{align*}

From this identity for  local representatives of the  stochastic process ${\tilde{\xi}_{\tilde\Sigma}}(t)$ it follows that
  (\ref{girs}) can be rewritten as 
\begin{eqnarray}
\!\!\!\!\!\!\!\!\!\frac{d{\mu}^{{\xi}_{\tilde\Sigma}}}
{d{\mu}^{\tilde{\xi}_{\tilde\Sigma}}}
({\tilde{\xi}_{\tilde\Sigma}}(t))
&=&\left(\frac{exp(\sigma({\tilde{\xi}_{\tilde\Sigma}}(t)}{exp(\sigma({\tilde{\xi}_{\tilde\Sigma}}(t_a)}\right)^{1/4}
\nonumber\\
&&\times\exp\Bigl\{-\frac18\mu^2\kappa\int^{t}_{t_a}\bigl(\tilde\triangle_{\tilde{\Sigma}}\sigma +\frac14<\partial\sigma,\partial \sigma>_{\tilde{\Sigma}}\bigr)ds\Bigr\},
\label{girs_result}
\end{eqnarray}
where by $\tilde\triangle_{\tilde{\Sigma}}\sigma $  we  denote the expression in the square brackets on the right-hand side of the previous formula. We see that in addition to the standard differential expression used to define the Laplace-Beltrami operator, it includes two more terms: $2j^{ A}_{\Roman 1}\sigma_A$ and $2j^{ a}_{\Roman 1 }\sigma_a$.
It can be shown that the sum of these terms is equal to zero.

Using the representations (\ref{j_A_deriv_N}) and (\ref{j_a_deriv_N}) for $j^{ A}_{\Roman 1}$ and $j^{ a}_{\Roman 1 }$, we rewrite this sum as follows:
$$h^{BM}(N^A_{B,M}\sigma_A+N^a_{B,M}\sigma_a)+(K^A_{\alpha}\sigma_A+K^a_{\alpha}\sigma_a)\,\Lambda^{\alpha}_C\,h^{\tilde B \tilde M}\,{}^{ \mathrm  H}{\tilde \Gamma}^{ C}_{\tilde B\tilde M}.$$

To prove that the expression in the second bracket of the above expression is equal to zero, one must first apply the operator $K^A_{\alpha} \partial_A+K^a_{\alpha} \partial_a$ to the metric $d_{\mu \nu}=K^C_{\mu}G_{CD}K^D_{\nu}+K^p_{\mu}G_{pq}K^q_{\nu}$  and then transform the resulting expression using Killing's relations from  Appendix. This leads to
\[
 K^A_{\alpha} \partial_Ad_{\mu\nu}+K^a_{\alpha} \partial_ad_{\mu\nu}=d_{\nu\sigma}c^{\sigma}_{\alpha\mu}+d_{\mu\sigma}c^{\sigma}_{\alpha\nu}.
\]
 By multiplying both sides of this equality by $d^{\mu\nu}$, we can conclude that the right-hand side of the equality is equal to zero for semisimple Lie groups.

The expression in the first bracket of the  equality under study is also equal to zero. This can be shown using the following representations:  $$N^A_{B,M}=-K^A_{\alpha,M}{\Lambda}^{\alpha}_B-K^A_{\alpha}{\Lambda}^{\alpha}_{B,M}\;\; \mathrm{and} \;\; N^a_{B,M}=-K^a_{\alpha}{\Lambda}^{\alpha}_{B,M},$$ 
and taking into account that
$h^{BM}{\Lambda}^{\alpha}_B=0$ and $K^A_{\alpha}\sigma_A+K^a_{\alpha}\sigma_a=0$.

The global integral relation between the Green's functions for the  reduction onto the zero-momentum level is given by
\begin{eqnarray*}
 &&d_b^{-1/4}d_a^{-1/4}G_{\tilde{\Sigma}}(\pi'_{(\tilde{\Sigma})}(p_b,v_b),t_b;
 \pi'_{(\tilde{\Sigma})}(p_a,v_a),t_a)\nonumber\\
 &&\;\;\;\;\;\;\;\;\;\;\;\;\;\;\;\;\;\;\;\;\;\;\;\;\displaystyle=\int _{\cal G}G_{\tilde{\cal P}}(p_b\theta,v_b\theta,t_b;
p_a,v_a,t_a) d\mu(\theta) , 
\end{eqnarray*}
where  $d_b$ and $d_a$ are the values of the $\det (d_{\alpha\beta})$ taken at the points $\pi'_{(\tilde{\Sigma})}(p_b,v_b)$ and $\pi'_{(\tilde{\Sigma})}(p_a,v_a)$.
The Green's function $G_{\tilde{\Sigma}}$ is presented by the following path integral:  
\begin{eqnarray*}
 &&G_{\tilde{\Sigma}}(\pi'_{(\tilde{\Sigma})}(p_b,v_b),t_b;
 \pi'_{(\tilde{\Sigma})}(p_a,v_a),t_a)\\
 &&=\int\limits_{{\tilde{\xi}_{\tilde \Sigma} (t_a)=\pi'_{(\tilde\Sigma)} (p_a,v_a)}\atop  
{\tilde{\xi}_{\tilde \Sigma} (t_b)=\pi'_{(\tilde\Sigma)} (p_b,v_b)}} d{\mu}^{\tilde{\xi}_{\tilde \Sigma}}
\exp \Bigl\{\frac 1{\mu ^2\kappa
m}\int_{t_a}^{t_b}\Bigl[\tilde{V}(\tilde{\xi}_{\tilde \Sigma}(u))+J(\sigma (\tilde{\xi}_{\tilde \Sigma}(u)))\Bigr]du\Bigr\},
\nonumber\\
\end{eqnarray*}
where  $J$ is the integrand in the reduction Jacobian (\ref{girs_result}):
\begin{equation}
J=-\frac18\mu^2\kappa\bigl(\triangle_{\tilde{\Sigma}}^{{\scriptscriptstyle \rm H}}\sigma +\frac14<\partial\sigma,\partial \sigma>_{\tilde{\Sigma}}\bigr).
\label{jacobian_reduct}
\end{equation}
In this expression, 
$$\triangle_{\tilde{\Sigma}}^{{\scriptscriptstyle \rm H}}\sigma=
h^{AB}\sigma_{AB}+2h^{Ab}\sigma_{Ab}
+h^{ab}\sigma_{ab}
-h^{\tilde B \tilde M}\,{}^{ \mathrm  H}{\tilde \Gamma}^{ A}_{\tilde B\tilde M}\sigma_A
-h^{\tilde B \tilde M}\,{}^{ \mathrm  H}{\tilde \Gamma}^{ a}_{\tilde B\tilde M}\sigma_a.$$

The global semigroup  determined by the Green's function $G_{\tilde{\Sigma}}$ acts in the Hilbert  space of the scalar functions on $\tilde{\Sigma}$ with the following scalar product $(\psi_1,\psi_2)=\int \psi_1\, \psi_2\,\,dv_{\tilde{\Sigma}}.$

The Green's function $G_{\tilde{\Sigma}}$ satisfies the forward Kolmogorov equation with the operator
\[
\hat{H}_{\kappa}=
\frac{\hbar \kappa}{2m}\tilde\triangle _{\tilde{\Sigma}}-\frac{\hbar \kappa}{8m}\Bigl[\triangle_{\tilde{\Sigma}}^{{\scriptscriptstyle \rm H}}\sigma +\frac14<\partial\sigma,\partial \sigma>_{\tilde{\Sigma}}\Bigr]+\frac{1}{\hbar \kappa}\tilde V.
\]
Note that at $\kappa =i$ the forward Kolmogorov equation becomes
the Schr\"odinger equation with the Hamilton operator 
$\hat H=-\frac{\hbar}{\kappa}{\hat H}_{\kappa}\bigl|_{\kappa =i}$.

Note also that the geometric properties of the operator $\tilde\triangle _{\tilde{\Sigma}}$,
$$\tilde\triangle_{\tilde{\Sigma}} =\triangle_{\tilde{\Sigma}}^{{\scriptscriptstyle \rm H}}+2j^{ A}_{\Roman 1}\,\partial_A+2j^{ a}_{\Roman 1}\,\partial_a,$$
as well as  the fact that this operator could be transformed into the Laplace-Beltrami operator given on the orbit space $\tilde{\mathcal M}$ with the Riemannian metric  (\ref{metric_h_ija}), if in charts of the reduced manifold it were possible to find   invariant local coordinates $(x^i,\tilde f^a)$, such that  $\chi^{\alpha}(Q^{\ast}(x^i))\equiv0$,  lead us to conlusion that the global semigroup  with the kernel $G_{\tilde{\Sigma}}$    can be used to describe the diffusion  of  interacting scalar particles on the orbit space $\tilde{\mathcal M}$ of the principal fiber bundle ${\rm P}(\tilde{\mathcal M},\mathcal G)$.

\section{Conclusion}
In this article we have considered the reduction procedure in the Wiener-type path integral which represents  the ``quantum evolution'' of a special finite-dimensional mechanical system with symmetry, consisting of two interacting scalar particles. 
The choice of such a system for study is due to its geometric properties, similar to the properties of those gauge theories that are used to describe the interaction of Yang-Mills fields with scalar fields.

Note that in order to apply the standard approach to the  quantization of  gauge theories, the original gauge fields must first be rewritten in terms of the constrained  variables.
In accordance with this, in the principal fiber bundle associated with our problem, a special system of local coordinates (including dependent coordinates) was introduced.
The  coordinates of this coordinate system are defined using the local surface  $\tilde{\Sigma}$ (a submanifold of the total space of the principal fiber bundle $\rm P(\tilde{\mathcal M},\mathcal G)$). And the reduced motion on the orbit space $\tilde{\mathcal M}$ is locally described in terms of these coordinates.

The reduction procedure performed in the article leads to the integral relation (\ref{green_funk_relat}) between  Green's functions
$G^{\lambda}_{mn}$ (the kernel of a reduced semigroup acting in the space of sections $\Gamma ({{\tilde \Sigma}},V^{*})$ for the general case of reduction onto the nonzero momentum level, $\lambda \neq 0$) and  $G_{\tilde{\cal P}}$, which is the kernel of the original semigroup. The representation of $G^{\lambda}_{mn}$ in terms of the path integral is given by the expression (\ref{path_int_G_mn}).

The   result obtained essentially depends on what assumption about the submanifold ${\tilde{\Sigma}}$ was made. Provided that   the submanifold ${\tilde{\Sigma}}$  is a global section of the principal fiber bundle $\rm P(\tilde{\mathcal M},\mathcal G)$ (which is thus a trivial principal bundle isomorphic to the trivial bundle   $\rm P(\tilde{\Sigma},\mathcal G$), then in this case it follows that the dependent coordinates $Q^{\ast A}$ (and the coordinates $\tilde f^a$)
can be thought of as global variables defined on ${\tilde{\Sigma}}$.
This case is analogous to what is usually assumed in gauge theories, when the evolution of the reduced system on the orbit space of  the gauge group is described in terms of variables with constraints.

However, in the most  cases, the principal fiber  bundles of gauge theories are nontrivial. Therefore, using this method, we can actually describe the `quantum' dynamics on the orbit space  only locally. Note that the problem of transition from a local description of the reduced  dynamics given in terms of  dependent coordinates to a global description is not solved yet. 

The  main result of  the article is the obtained expression (\ref{jacobian_reduct}) for the integrand $J$  in the path integral reduction Jacobian   for  the case of reduction onto a  zero-momentum level. In particular, it is shown that $J$ is generated by the projection of the  mean curvature vector field of the orbit onto the submanifold ${\tilde{\Sigma}}$. 

In addition, we note that the resulting $J$ is a generalization to interacting dynamical systems with symmetry of a similar expression obtained by J. Lott in \cite{Lott} while studying the quantum potential in pure Yang-Mills theory.

The Girsanov-Cameron-Martin transformation can also be performed in the path integral (\ref{path_int_G_mn}).
In this case, the exponential with the integral of $J$, defined by the formula (\ref{jacobian_reduct}), will be the diagonal part of the reduction Jacobian.

\appendix
\section*{Appendix }
\section*{Relationship between the Christoffel symbols}
\setcounter{equation}{0}
\def\theequation{A.\arabic{equation}}

\begin{eqnarray*}
 &&{\Gamma}^a_{kn}=\tilde h^{am}{\Gamma}_{knm}+\tilde h^{ab}{\Gamma}_{knb}
\nonumber\\
&&{\Gamma}^a_{kb}=\tilde h^{an}{\Gamma}_{kbn}+\tilde h^{ac}{\Gamma}_{kbc}
\nonumber\\
 &&{\Gamma}^a_{bm}=\tilde h^{an}{\Gamma}_{bmn}+\tilde h^{ac}{\Gamma}_{bmc}
\nonumber\\
 &&{\Gamma}^a_{bc}=\tilde h^{an}{\Gamma}_{bcn}+\tilde h^{ad}{\Gamma}_{bcd}
\nonumber\\
&&{}
\nonumber\\
&&{\Gamma}_{knb}={}^{ \mathrm  H}{\tilde \Gamma}_{BMb}\,Q^{\ast B}_kQ^{\ast M}_n+{\tilde G}^H_{Ab}Q^{\ast A}_{kn}
\nonumber\\
 &&{\Gamma}_{kbn}={}^{ \mathrm  H}{\tilde \Gamma}_{AbB}\,Q^{\ast A}_kQ^{\ast B}_n
\nonumber\\
 &&{\Gamma}_{kbc}={}^{ \mathrm  H}{\tilde \Gamma}_{Abc}\,Q^{\ast A}_k
\nonumber\\
 &&{\Gamma}_{bmn}={}^{ \mathrm  H}{\tilde \Gamma}_{bAB}\,Q^{\ast A}_mQ^{\ast B}_n
\nonumber\\
 &&{\Gamma}_{bmc}={}^{ \mathrm  H}{\tilde \Gamma}_{bAc}\,Q^{\ast A}_m
\nonumber\\
&&{}
\nonumber\\
 &&{}^{ \mathrm  H}{\tilde \Gamma}_{AbB}={\tilde G}^H_{\tilde RB}{}^{ \mathrm  H}{\tilde \Gamma}^{\tilde R}_{Ab}
\nonumber\\
 &&{}^{ \mathrm  H}{\tilde \Gamma}_{Abc}={\tilde G}^H_{\tilde Rc}{}^{ \mathrm  H}{\tilde \Gamma}^{\tilde R}_{Ab}
\nonumber\\
 &&{}^{ \mathrm  H}{\tilde \Gamma}_{bAB}={\tilde G}^H_{\tilde RB}{}^{ \mathrm  H}{\tilde \Gamma}^{\tilde R}_{bA}
\nonumber\\
 &&{}^{ \mathrm  H}{\tilde \Gamma}_{bAc}={\tilde G}^H_{\tilde Rc}{}^{ \mathrm  H}{\tilde \Gamma}^{\tilde R}_{bA}
\nonumber\\
 &&{}
 \nonumber\\
 &&{\Gamma}^a_{kn}=G^{EF}N^a_E{\tilde G}^H_{\tilde RF}{}^{ \mathrm  H}{\tilde \Gamma}^{\tilde R}_{BM}Q^{\ast B}_kQ^{\ast M}_n+G^{ab}{\tilde G}^H_{\tilde Rb}{}^{ \mathrm  H}{\tilde \Gamma}^{\tilde R}_{BM}Q^{\ast B}_kQ^{\ast M}_n+N^a_EQ^{\ast A}_{kn}
\nonumber\\
 &&{\Gamma}^a_{kb}=N^a_{\tilde R}\,{\tilde \Gamma}^{\tilde R}_{Ab}\,Q^{\ast A}_k
\nonumber\\
 &&{\Gamma}^a_{bm}=N^a_{\tilde R}\,{\tilde \Gamma}^{\tilde R}_{bA}\,Q^{\ast A}_m
\nonumber\\
 &&{\Gamma}^a_{bc}=N^a_{\tilde R}\,{\tilde \Gamma}^{\tilde R}_{bc}
\nonumber\\
\end{eqnarray*}

{\bf Killing relations}
\begin{align*}
 &\;\;\;\;\,K^A_{\alpha}G_{CD,A}=-G_{CR}K^R_{\alpha,D}-G_{RD}K^R_{\alpha,C},\\
& -G_{qb'}K^{b'}_{\alpha, a}-G_{ab'}K^{b'}_{\alpha,q}=0.
\end{align*}

{\bf Properties of the horizontal metric ${\tilde G}^H_{\tilde A\tilde B}$}
\begin{eqnarray*}
&&N^T_F{\tilde G}^H_{\tilde RT}+N^b_F{\tilde G}^H_{\tilde Rb}={\tilde G}^H_{\tilde RF},\\
&&G^{ab}{\tilde G}^H_{Ab}=\tilde{\Pi}^a_A,\;\;\;G^{EF}{\tilde G}^H_{AF}=\tilde{\Pi}^E_A,\;\;\;N^a_E\tilde{\Pi}^E_A+\tilde{\Pi}^a_A=N^a_A.
\end{eqnarray*}


\begin{thebibliography}{**}
\bibitem{Faddeev} 
L. D. Faddeev, The Feynman integral for singular Lagrangians, Theoretical and Mathematical Physics, {\bf 1} 1 (1969), 
{\it Teor. i Mat. Fyz.} {\bf 1} 3 (1969)  (in Russian);\\
L. D. Faddeev, V. N. Popov, Feynman diagrams for the Yang-Mills field,  Phys. Lett. B 
{\bf 25} 29 (1967).

\bibitem{Storchak_2}
S. N. Storchak, 
 Dependent coordinates in path integral measure factorization, 
J. Phys. A: Math. Gen. 
 {\bf 37} 7019 (2004),
(IHEP Preprint 2000-54, Protvino, 2000), arXiv: math-ph/0311038.

\bibitem{Storchak_2009}
S. N. Storchak, 
 On the geometrical representation of the path integral reduction Jacobian: The case of dependent variables in description of reduced motion,
J. Geometry and Physics,
 {\bf 59} 1155 (2009).

\bibitem{Mitter-Viallet}
P. K. Mitter and C. M. Viallet, On the bundle of connections and the gauge orbit manifold in Yang-Mills theory, Commun. Math. Phys. 
{\bf 79} 457 (1981).

\bibitem{Storchak_1} 
S. N. Storchak, 
 Path integrals on manifold with group action, 
J. Phys. A: Math. Gen.
 {\bf 34} 9329 (2001), (
IHEP Preprint 96-110, Protvino, 1996).

\bibitem{Storchak_98}  
S. N. Storchak, 
{\it Bogolubov transformation in path integrals 
on manifold with a group action}
IHEP Preprint 98--1, Protvino, 1998.

\bibitem{Dalecky_1}
Ya. I. Belopol'skaya  and Yu. L. Daletskii, Itô equations and differential geometry,
Russ. Math. Surv. {\bf 37} no 3,   109 (1982);\\
Yu. L. Daletskii, Stochastic differential geometry,
 Russ. Math. Surv.  {\bf 38} no 3,  97 (1983). 

\bibitem{Dalecky_2}
Ya. I. Belopol'skaya  and Yu. L. Daletskii, 
{\it Stochastic equations and differential geometry} 
(Kluwer Academic Publishers,1990).

\bibitem{Elworthy}
 K. David Elworthy, Xue-Mei Li, Yves LeJan, {\it The Geometry of Filtering} (Birkhauser, 2010). 

 \bibitem{Storchak_2019} 
S. N. Storchak, Path integrals on a manifold that is a product of the total space of the principal fiber bundle and the vector space,
arXiv:1912.13124. 

\bibitem{Storchak_2020} 
S. N. Storchak, On the geometric representation of the path integral reduction Jacobian for a mechanical system with symmetry given on a manifold that is a product of the total space of the principal fiber bundle and the vector space,
arXiv:2007.04397. 

\bibitem{AbrMarsd}
R. Abraham, J. E. Marsden, {\it Foundation of Mechanics, 2nd Ed.}
(Addison-Wesley Redwood City, 1985).

\bibitem{Creutz}
M. Creutz, I. J. Muzinich, and T. N. Tudron, Gauge fixing and canonical quantization, 
 Phys. Rev. D {\bf 19} no. 2, 531 (1979).

\bibitem{Huffel-Kelnhofer}
H. H\"{u}ffel and G. Kelnhofer,  QED revisited: proving equivalence between path integral and stochastic quantization,  Phys. Lett. B   {\bf 588} 145 (2004).

\bibitem{Kelnhofer_2}
H. H\"uffel and G. Kelnhofer,  Generalized Stochastic Quantization of Yang-Mills Theory,
Ann. of Phys. {\bf 270}  231 (1998).

\bibitem{Lipcer}
R. S. Lipster and A. N. Shiryayev, Statistics of Random
Processes, Vols. {\Roman 1} and {\Roman 2} (Springer--Verlag:
Berlin, Heidelberg, New York, 1977). 

\bibitem{Pugachev}
V. S. Pugachev and I. N. Sinitsyn, Stochastic Differential Systems,
2nd Edition
(Moscow, Nauka, 1990) (in Russian).


\bibitem{Dalmulti}
Yu. L. Dalecky, N. I. Teterina, Multiplicative stochastic integrals,
 Usp. Mat. Nauk {\bf 27} no 2, 167
(1972) (in Russian);\\
Yu. L. Dalecky, Multiplicative operators of diffusion processes and differential equations in sections of vector bundles,  Usp. Mat. Nauk {\bf 30} no 2, 209 (1975)  (in Russian).

\bibitem{Stroock}
D. W. Stroock, On certain systems of parabolic equations, Com. Pure Appl. Math. 
{\bf 23} 447 (1970).

\bibitem{Lewis}
J. T. Lewis, Brownian motion on a submanifold of Euclidean space, Bull. London Math. Soc. {\bf 18} 616 (1986).

\bibitem{Betounes}
D. E. Betounes, Mean-curvature normal and dual-string models,  Phys. Rev. D {\bf  33} 3634 (1980).

\bibitem{Lott}
J. Lott, The Yang-Mills Collective-Coordinate Potential, Commun. Math. Phys. {\bf 95}, 289 (1984).

\end{thebibliography}
\end{document}